\documentclass[11pt, oneside]{article}   	% use "amsart" instead of "article" for AMSLaTeX format
\usepackage{geometry}                		% See geometry.pdf to learn the layout options. There are lots.
\usepackage{graphicx}				% Use pdf, png, jpg, or eps§ with pdflatex; use eps in DVI mode
\usepackage{amssymb}
\usepackage{tikz}
\usepackage{latexsym, amsmath, mathrsfs, graphicx}
\usepackage[utf8]{inputenc}
\newcommand{\beq}{\begin{eqnarray}}
\newcommand{\eeq}{\end{eqnarray}}
\usepackage{authblk}
\usepackage[super,compress]{cite}
\usepackage{subfig}
\usepackage{graphics}

\newcommand{\qbar}{{\bar q}}

\newcommand{\be}{\begin{equation}}
\newcommand{\ee}{\end{equation}}
\newcommand{\bea}{\begin{eqnarray}}
\newcommand{\eea}{\end{eqnarray}}
\newcommand{\Ei}{\text{Ei}}

\title{TMD gluon distributions at small x in the CGC theory}
\author{Elena Petreska
\footnote{petreska@nikhef.nl}}
\affil{Department of Physics and Astronomy, VU University Amsterdam, De Boelelaan 1081,\\
NL-1081 HV Amsterdam,
The Netherlands
 }
 \affil{Nikhef, Science Park 105, \\
NL-1098 XG Amsterdam, The Netherlands}
\date{}							% Activate to display a given date or no date

\begin{document}

\hspace{5.3in} \mbox{Nikhef 2018-016} 

\let\newpage\relax
\maketitle

\begin{abstract}
We review recent progress in the description of unpolarized transverse-momentum-dependent (TMD) gluon distributions at small $x$ in the color glass condensate (CGC) effective theory. We discuss the origin of the non-universality of TMD gluon distributions in the TMD factorization framework and in the CGC theory and the equivalence of the two approaches in their overlapping domain of validity. We show some applications of this equivalence, including recent results on the behavior of TMD gluon distributions at small $x$, and on the study of gluon saturation. We discuss recent advances in the unification of the TMD evolution and the non-linear small-$x$ evolution of gluon distributions.

\end{abstract}
%\section{}

%\keywords{TMD factorization; CGC; TMDs at small $x$, gluon saturation.}

\section{Introduction}

The dynamical structure of hadrons and nuclei can be studied via parton distribution functions (PDFs) which describe how their constituent particles (the partons) are distributed in one-dimensional (longitudinal) momentum at a given resolution scale. The PDFs are non-perturbative objects that one can extract from experiments, a procedure based on collinear factorization formulas which separate the parton densities from the perturbative short-distance part of the collision. Due to the universality of the parton distribution functions, they can be measured in one experiment and then used in another scattering process at a different resolution scale. The evolution equations that describe how the densities change with a change in the resolution scale can be derived within the perturbative framework of Quantum Chromodynamics (QCD) and are called the Dokshitzer-Gribov-Lipatov-Altarelli-Parisi (DGLAP) equations~\cite{1,2,3,4}. For some recent sets of PDFs extracted from a global analysis of experimental data see for example Refs.~\cite{Ball:2017nwa,Dulat:2015mca,Harland-Lang:2014zoa,Butterworth:2015oua}.

The collinear factorization approach with PDFs can be used for studying the most inclusive scattering processes at high energy and when the momentum transfer in the collision is large. For less inclusive observables one can become sensitive to the transverse momenta of the partons and the three-dimensional momentum structure of the participants in the collision becomes important. The quantities that describe the structure of hadrons and nuclei in three-dimensional momentum space (in longitudinal and transverse momentum), including polarization degrees of freedom, are called transverse-momentum-dependent parton distribution functions (TMD PDFs or TMDs for short). The TMDs are also non-perturbative quantities that can be extracted from experiments with the help of TMD factorization formulas~\cite{Collins:CUP2011} and which obey QCD evolution equations (called TMD evolution) that are a generalization of the evolution equations for PDFs~\cite{Collins:CUP2011,Collins:1981uw,Collins:1981va,Collins:1982wa,Collins:1985}. The TMD factorization formulas that have been derived for 
%positron-electron annihilation into two hadrons~\cite{Collins:1981uk}, 
semi-inclusive deep-inelastic scattering (SIDIS)~\cite{Ji:2004wu} and Drell-Yan and Z boson production in proton-proton (pp) collisions~\cite{Collins:CUP2011} have been recently used to extract the unpolarized quark TMD from a global data analysis for the first time~\cite{Bacchetta:2017gcc}. 

Unlike collinear parton densities, TMD distributions are not universal, their operator definition depends on the process under consideration~\cite{Collins:1983pk,Brodsky:2002cx,Collins:2002kn,12,Bomhof:2004aw,Bomhof:2006dp,Boer:1999si}, and TMD factorization with universal distributions is not always possible. The non-universality of TMDs finds its origin in the color structure of the short-distance part of the collision, it is a consequence of the fundamental properties of QCD and has attracted much scientific attention recently. The process dependence of TMDs is one of the main topics in this review.

A particularly interesting kinematic region to study TMDs and their broken universality is the small-$x$ saturation limit, where $x$ is the longitudinal-momentum fraction of the parton with respect to its parent hadron. 
%The evolution of parton densities with decreasing $x$ at fixed resolution scale is given with the linear Balitsky-Fadin-Kuraev-Lipatov (BFKL) equations~\cite{13,14}. 
At high energy, the particle content of ultra-relativistic protons and nuclei is dominated by gluons whose density increases as $x$ decreases according to the linear Balitsky-Fadin-Kuraev-Lipatov (BFKL) evolution equations~\cite{13,14}. At a given point in the small-$x$ evolution, when the proton or nucleus becomes a very dense system of gluons, the rise of the gluon density becomes affected by non-linear effects which slow down its growth~\cite{15,16,17,18,19,20,21,22}. An effective theory of QCD that describes the dynamics of partons in high density systems and the phenomenon of gluon saturation at small $x$ is the Color Glass Condensate (CGC) theory~\cite{19,20,21,22}. The transition between the linear and non-linear regimes in the CGC framework is characterized with a momentum scale called saturation momentum ($Q_s$). Because of this newly-generated intrinsic momentum scale in the problem, a proper description of the gluon dynamics naturally requires knowledge of the transverse momentum of the gluons (compared to $Q_s$) and of their distributions in all three momentum components. These are usually referred to as $k_t$-dependent or unintegrated gluon distributions (UGDs). The non-universality of UGDs at small $x$ was observed in Ref.~\cite{Kharzeev:2003wz} where the authors distinguished between two gluon distributions, the dipole and the Weisz\"acker-Williams (WW) distribution. The process dependence of UGDs as a reason for their non-universality was realized in Refs.~\cite{Dominguez:2010xd,Dominguez:2011wm}.

%There are two fundamental distributions in the CGC theory, the dipole gluon distribution associated with a quark-antiquark pair scattering off the dense gluon field and which appears in several high-energy processes, and the Weisz\"acker-Williams distribution which gives the number density of gluons in the hadron or nucleus.

The CGC is a universal theory in the small-$x$ saturation limit, natural for any high-energy, ultra-relativistic proton or nucleus, independent of the particular scattering process.~\cite{22} The CGC theory has been successfully applied for describing and predicting experimental data in deep-inelastic scattering (DIS), pp, proton-nucleus (pA) and nucleus-nucleus (AA) collisions, for a variety of processes and observables. Cross sections derived in the CGC theory involve products of Wilson lines averaged over the color charge distribution in the proton or the nucleus, which is a universal object independent of the particular process. Its evolution with $x$ in the non-linear regime is given by the Jalilian-Marian-Iancu-McLerran-Weigert-Leonidov-Kovner (JIMWLK) renormalization group equation~\cite{JIMWLK1,JIMWLK2,JIMWLK3,JIMWLK4,JIMWLK5,JIMWLK6} or the Balitsky-Kovchegov~\cite{balitsky1,balitsky2,balitsky3,Kovchegov} (BK) evolution equation. 

The natural question arises whether some form of universality of the TMD distributions can be restored within the CGC theory at small $x$. The first step to answering this is proving equivalence between the TMD factorization approach to high-energy scatterings and the CGC theory at small $x$ in their overlapping domain of validity. For physical situations where both formalisms can be applied, \textit{i.e.}\ for processes at small $x$, such that the CGC theory is applicable, and which involve ordering of momentum scales, such that TMD factorization can be derived, they should yield the same results. To outline the physical picture, we will review a set of processes for which an explicit equivalence between the two approaches has been established. We will consider dijet production in pp and pA collisions~\cite{Dominguez:2011wm,Dominguez:2010xd,Marquet:2016cgx}, dijet production in DIS~\cite{Dominguez:2011wm,Dominguez:2010xd}, direct-photon jet production in pp and pA collisions~\cite{Dominguez:2011wm,Dominguez:2010xd} and Higgs boson production in pp and pA collisions.~\cite{Sun:2011iw,Boer:2011kf,Schafer:2012yx}. We will discuss the case of dijet production in pp and pA collisions in detail and explain the non-universality of TMDs for this particular situation. The analysis can be extended to the rest of the processes and they will be briefly reviewed.

%It was shown that the TMD distributions in the small-$x$ limit can be identified with the dipole distribution and the Weisz\"acker-Williams distribution from the CGC theory, or with a convolution of both. 

The process of dijet production in pA collisions at forward rapidity (in the direction of the proton) is particularly interesting as it can be studied with both TMD and CGC frameworks, so it is suitable for their comparison, and in addition, it can be used to study saturation effects in the nucleus. The magnitude of the total transverse momentum (the momentum imbalance) of the jet pair, $k_t$, probes the transverse momenta of the partons in the nucleus. When $k_t$ is much smaller than the typical transverse momentum of the individual jets, $P_t$, \textit{i.e.}\ in the correlation limit, the cross section can be written in a factorized form to leading order in $k_t/P_t$. The cross section involves on-shell matrix elements representing the short-distance part of the process (the $2\to 2$ parton scattering), PDFs describing the proton and TMD gluon distributions describing the nucleus~\cite{Dominguez:2011wm,Kotko:2015ura}.

On the other hand, the cross section for this process, in the small-$x$ limit, can be independently derived using CGC methods without imposing restrictions on the ordering between the momentum imbalance and the momentum of the jets and it is valid for any $k_t$ values between $Q_s$ and $P_t$~\cite{Dominguez:2011wm,Marquet:2007vb,Iancu:2013dta}. Two jets produced in the fragmentation region of the proton probe the proton wave function at large values of $x$ (the proton is dilute) while on the nucleus side they probe the small-$x$ gluons (the nucleus is dense). The momentum imbalance $k_t$ probes the transverse momenta of the small-$x$ gluons in the nucleus. For this type of asymmetric collisions one usually applies the hybrid approach~\cite{Dumitru:2005gt} in which the dilute proton is described with collinear PDFs, while the opposite moving nucleus is represented by a dense gluon field. This asymmetric hybrid approach is also applicable in pp collisions when the jet production is sufficiently forward. The multiple rescatterings of the incoming partons from the proton off the gluon field in the nucleus are resummed into Wilson lines. The cross section is then expressed in terms of correlators of multi-point Wilson lines.

In order to compare the results for the cross sections derived using TMD factorization and CGC methods, one needs to apply appropriate limits on both sides and study the results in the common region of applicability of the two theories. Refs.~\cite{Dominguez:2010xd,Dominguez:2011wm} showed that the small-$x$ limit of the TMD factorized cross section is equivalent to the correlation limit ($k_t \ll P_t$) of the CGC result. At small $x$ each of the gluon TMDs can be defined as an UGD in the CGC framework, {\it i.e.}\ as an operator of Wilson lines averaged over the distribution of color charges in the proton or nucleus~\cite{Dominguez:2011wm,Marquet:2016cgx}. The CGC distribution of color charges is universal among different processes, hence in this sense restoring universality of gluon TMDs at small $x$.

Once the equivalence between the TMD and CGC approaches in their overlapping domain is proven, we can use CGC methods to study the properties of TMD distributions at small $x$. We will devote part of this review to the small-$x$ behavior of gluon TMDs that follows from results in the CGC theory. The TMD factorization formula for forward dijet production in pA collisions for a finite number of colors ($N_c$) involves eight different unpolarized gluon TMDs~\cite{Kotko:2015ura}, including all that have been identified in other processes, which allows us to study the complete set of TMDs appearing in cross sections so far. We will discuss results for the gluon distributions in the Golec-Biernat-Wusthoff (GBW) model~\cite{GolecBiernat:1998js}, in the McLerran-Venugopalan (MV) model~\cite{17,18}, and their JIMWLK small-$x$ evolution. The main conclusions will be that all TMDs have the same behavior (they are universal) at large $k_t$, while they behave differently at small $k_t$ (though the differences are under control in the CGC) and that after some evolution they become functions of $k_t/Q_s(x)$ only (they reach geometric scaling).~\cite{Marquet:2016cgx}

We will also review recent progress made in unifying the TMD evolution and the non-linear small-$x$ evolution of gluon TMDs. As discussed above, TMD factorization formulas can be derived when there exists an ordering of momentum scales, {\it i.e.}\ when $k_t \ll Q$ with $k_t$ a soft scale and $Q$ the hard scale relevant in the process. Because of this ordering, large contributions of the type $ \ln^2 Q^2/k_t^2$ (called Sudakov double logarithms~\cite{Sudakov:1954sw}) become important and powers of $ \alpha_s\ln^2 Q^2/k_t^2$ need to be resummed. The TMD evolution (or Collins-Soper evolution)~\cite{Collins:1981va} includes the resummation of these double logarithms (called Collins-Soper-Sterman resummation)~\cite{Collins:1985}. On the other hand, the small-$x$ evolution (JIMWLK or BK) resums terms of the type $\alpha_s \ln\, 1/x$. It is important to have a unified framework that resums both types of large logarithms simultaneously and consistently. The type of large logarithms that dominate would be determined by the kinematics of the particular process. We will review some recent investigations along these lines~\cite{Mueller:2012uf,Mueller:2013wwa,Balitsky:2015qba,Balitsky:2016dgz,Xiao:2017yya}.

The final topic that we will address in this review is an improved TMD factorization model with an extended range of validity, which can be used to study saturation effects in forward dijet production in pA collisions and in ultra-peripheral heavy ion collisions (UPC). The TMD factorization formula for dijet production in pA collisions is valid in the correlation limit ($k_t \ll P_t$), it involves on-shell matrix elements and several TMD gluon distributions describing the nucleus. In the limit of large $k_t$ of the order of the hard scale ($k_t \sim P_t$) there exists a factorization with off-shell matrix elements and one unintegrated gluon distribution for the nucleus, called high energy factorization (HEF)~\cite{Catani:1990eg,Deak:2009xt}. Both of these formulas can be derived from perturbation theory, but away from their limits there is no established factorization in terms of short-distance matrix elements and long-distance parton distributions. The CGC cross section does capture both of these limits and the region in between but it is not a factorization formula in the usual sense of the word and is more complicated for phenomenological applications. Instead, one can use a model, an improved TMD factorization (ITMD) that is applicable for any value of $k_t$ between $Q_s$ and $P_t$ with off-shell matrix elements and several gluon TMDs.~\cite{Kotko:2015ura} The ITMD formula is an interpolation between the TMD and HEF cross sections and it reproduces both of them in the appropriate limits. Using the ITMD framework one can observe an onset of saturation effects in the region where $k_t$ approaches $Q_s$ in forward dijet production in pA collisions~\cite{vanHameren:2016ftb} and in UPC~\cite{Kotko:2017oxg}.

In the end, we want to point out that the review will concentrate only on unpolarized TMD gluon distributions. Significant progress along similar lines as discussed here has been achieved for polarized parton and/or hadron or nucleus (see {\it e.g.}\ some recent works, Refs.~\cite{Kovchegov:2015zha,Dumitru:2016jku,Boer:2016fqd,Boer:2016xqr,Szymanowski:2016mbq,Kovchegov:2016zex,Kovchegov:2016weo,Marquet:2017xwy,Boer:2017xpy,Kovchegov:2017lsr,Kovchegov:2017jxc}, and references therein for previous contributions).

The review is organized as follows. In section~\ref{sec:TMD} we recall the origin of the non-universality of gluon TMDs in the TMD factorization approach and discuss the derivation of an effective TMD factorization for a set of processes. In section~\ref{sec:CGC} we briefly review some main aspects of the CGC theory and outline the derivation of CGC cross sections. In section~\ref{sec:equiv} we show the equivalence between the small-$x$ limit of the effective TMD factorization and the correlation limit of the CGC cross section. In section~\ref{TMDsmall-x} we discuss main properties of the WW and dipole gluon TMDs. In section~\ref{sec:results} we review some recent results on the small-$x$ behavior of gluon TMDs and on gluon saturation based on the ITMD framework. In section~\ref{sec:evolution} we summarize recent results on the combination of the TMD evolution and non-linear small-$x$ evolution of gluon TMDs. In section~\ref{sec:conclusions} we conclude with final remarks.
%, while the processes of dijet production in DIS, photon-jet production in pp and pA collisions and Higgs boson production in pp and pA collisions will be briefly reviewed.

\section{Effective TMD Factorization and Non-Universality of TMDs}
\label{sec:TMD}

In this section we review a set of processes for which an effective TMD factorization has been derived and the equivalence with the CGC cross section has been shown. We study dijet production in pp and pA collisions in detail to demonstrate the non-universality of gluon TMDs.

\subsection{Forward dijet production in pA collisions}

In the most general case of dijet production in pp (pA) collisions, it is not possible to derive a factorized form of the cross section in terms of TMD distribution functions for the protons (for the proton and the nucleus).~\cite{Bomhof:2006dp,Boer:2003tx,Qiu:2007ar,Qiu:2007ey,Collins:2007nk,Rogers:2010dm,Xiao:2010sp} However, when both of the jets in pA collisions are produced at forward rapidity, such that one can neglect the transverse momentum of the partons in the proton, one can derive an effective TMD factorization formula with collinear PDFs for the proton and TMD distribution functions for the nucleus, in the back-to-back (correlation) limit~\cite{Dominguez:2011wm,Kotko:2015ura}. The same form of factorization holds for asymmetric pp collisions as well, with PDFs for the dilute, and TMDs for the dense proton. In the following, we will discuss pA collisions, having in mind that the same analysis can be also applied to pp collisions, although using a nucleus enhances the asymmetry.

We will first analyze the kinematics for inclusive production of two jets at forward rapidity in pA collisions. It is important to distinguish between different kinematic regions of the same process for which different theoretical frameworks are relevant. In particular, we will make a separation between the cases when the jets are produced almost back-to-back in the transverse plane (correlation limit) and when they are away from this  scenario. As discussed in the introduction, in the first limit one can apply the TMD factorization formula, while in the second case the HEF cross section is suitable. We will show below that the CGC cross section contains both of these limits.

\subsubsection{Kinematics for dijet production in pp and PA collisions}
\label{sub:Kinematics}

\begin{figure}
  \begin{center}
    \includegraphics[width=0.30\textwidth]{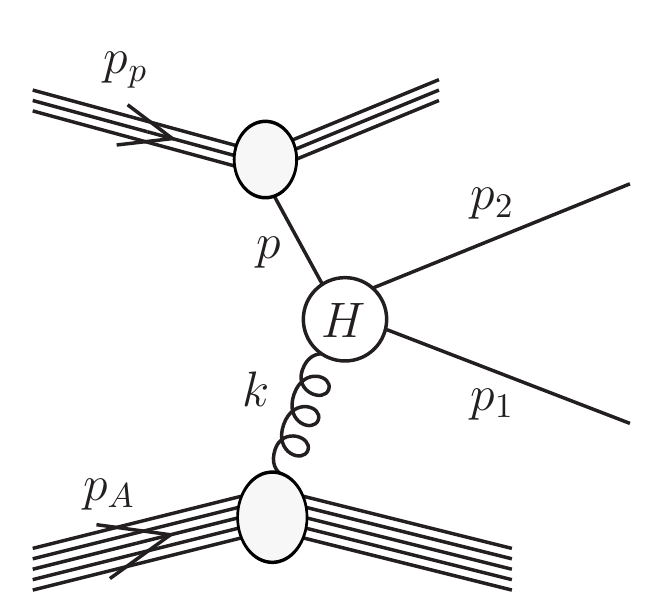}
  \end{center}
  \caption{Inclusive production of two jets in a pA collision. The circle $H$ represents
  hard scattering and the lines 
  connected to it represent partons, either quarks or gluons. Figure from Ref.~\protect\cite{Marquet:2016cgx}.
   }
  \label{fig:dijets-pA}
\end{figure}

We are interested in the process:
\begin{equation}
  p (p_p) + A (p_A) \to j_1 (p_1) + j_2 (p_2)+ X\,,
\end{equation}
where $p_p$ and $p_A$ are the four-momenta of the proton and the nucleus respectively, and $p_1$ and $p_2$ the four-momenta of the produced jets. The process is given schematically in Fig.~\ref{fig:dijets-pA}. The incoming momenta $p_p$ and $p_A$ only have longitudinal components; the proton is moving in the $+z$ direction and the nucleus in the $-z$ direction. We use light-cone coordinates defined as $v^\pm = (v^0\pm v^3)/\sqrt{2}$ and the notation $v^\mu=(v^+,v^-,{\vec {v}_t})$\footnote{We will also use the notation $v^\mu=(v^+,v^-,{\bf v})$ for position space coordinates.}. The four-momenta of the proton and the nucleus are $p_p = \sqrt{{s}/{2}}(1,0,\vec 0_t)$ and $p_A = \sqrt{{s}/{2}}(0,1,\vec 0_t)$ with $s$ the center of mass energy squared of the p+A system. The rapidities of the produced jets determine the values of $x$ that are probed in the proton and the nuclear wave functions. We denote with $x_1$ and $x_2$ the longitudinal-momentum fractions of the incoming partons from the proton and from the nucleus, respectively. Their values are given in terms of the rapidities $y_1$ and $y_2$ and the magnitudes of the transverse momenta $p_{1t}$ and $p_{2t}$ of the produced jets:
\begin{subequations}
\begin{align}
x_1 & = \frac{p_1^+ + p_2^+}{p_p^+}   = \frac{1}{\sqrt{s}} \left(p_{1t}\,
e^{y_1}+p_{2t} \,e^{y_2}\right)\,, \\
x_2 & = \frac{p_1^- + p_2^-}{p_A^-}   = \frac{1}{\sqrt{s}} \left(p_{1t} \,e^{-y_1}+p_{2t}\, e^{-y_2}\right)\,.
\end{align}
\end{subequations}
When the two jets are produced at forward rapidity, $y_1,y_2 \gg 1$, the proton is probed at large values of $x$ ($x_1 \sim 1$), while the nucleus is probed at small $x$ ($x_2\ll 1$). Therefore, we assume that the proton is a dilute system of partons away from the region where non-linear effects are relevant, while the nucleus is a dense system of gluons with typical transverse momentum of the order of $Q_s$ where saturation effects are important. As gluons dominate in the nuclear wave function we will always consider an incoming gluon from the nucleus and calculate the partonic channels $qg \to qg$, $gg\to q\bar{q}$ and $gg\to gg$.

The total transverse momentum of the produced jets is $\vec k_t \equiv \vec p_{1t} + \vec p_{2t}$ and its magnitude is:
\begin{equation}
  k_{t}^2 = (\vec p_{1t}+ \vec p_{2t})^2 = 
  p_{1t}^2 + p_{2t}^2 + 2\, p_{1t}\, p_{2t} \cos\Delta\phi\,,
  \label{eq:ktglue}
\end{equation}
where $\Delta\phi$ is the angle between the transverse momenta of the jets. The hard momenta $p_{1t}$ and $p_{2t}$ provide the large momentum scale, while $Q_s$ is the soft scale in the problem. With respect to the third momentum scale $k_t$, we will consider two limits. When the jets are produced almost back-to-back ($\Delta \phi \sim \pi$) the momentum imbalance is small: $k_t \ll p_{1t},p_{2t}$, and we will take it to be of the order of the soft scale $Q_s$: $k_t\sim Q_s$. Away from the back-to-back limit the total transverse momentum is large and of the order of the momenta of the jets: $k_t \sim p_{1t}, p_{2t} \gg Q_s$. 

In both of these limits we assume that the momentum imbalance of the jets comes from the transverse momentum of the gluons in the nucleus. This is justified because we can neglect the transverse momentum of the partons in the proton compared to the transverse momentum of the gluons in the nucleus. In the first limit of small $k_t$, soft momenta are probed in the proton and nuclear wave function. In the dilute proton, the transverse momenta of the partons are of the order of $\Lambda_{QCD}$, while the transverse momenta of the gluons in the nucleus are of order $Q_s$. For $Q_s \gg \Lambda_{QCD}$ we can assume that the momentum imbalance of the jets $k_t \sim Q_s$ is acquired from the transverse momentum of the gluons in the nucleus. In the second limit when $k_t$ is hard, the momentum imbalance probes the perturbatively large transverse momenta in the proton and in the nucleus. Again, because of the asymmetry of the problem, DGLAP evolution of the proton densities implies $1/k_t^2$ behavior at large $k_t$, while BFKL evolution of the nucleus implies $1/k_t$ behavior. Because of their faster fall-off, we again neglect the transverse momenta of the partons in the proton compared to the transverse momenta of the gluons in the nucleus. We can therefore assume that the momentum imbalance of the jets comes from the transverse momenta of the gluons and we use the notation $k$ for the four-momentum of the gluon from the nucleus, as in Fig.~\ref{fig:dijets-pA}. We are therefore not sensitive to the transverse-momentum structure of the proton and we describe it with collinear PDFs.

\subsection{Non-universality of gluon TMDs}

In this subsection we consider nearly back-to-back jets in the transverse plane such that there is an ordering of momentum scales, $k_t \ll P_t$, and effective TMD factorization can be derived. In the correlation limit one can consider only the leading order in $k_t / P_t$ of the cross section. In this limit the dependence on $k_t$ survives only in the TMD gluon distributions, while $k_t=0$ in the partonic subprocesses and the hard part is calculated with on-shell particles. 

As follows from the kinematical arguments discussed above, we describe the proton with collinear PDFs, $f_{a/p}(x_1)$, which give the probabiblity of finding a parton of type $a$ (a quark, an antiquark or a gluon) in the proton wave function with longitudinal-momentum fraction $x_1 = p^+/p^+_p$ at a given resolution scale. The evolution of the PDFs with the resolution scale is given by the DGLAP equations.

%\begin{figure}[t]
%\centering
%\begin{minipage}{.5\textwidth}
%  \centering
%  %\begin{center}
%    \includegraphics[width=0.95\textwidth]{qg2qg-diag.png}
%  %\end{center}
%  \caption{Diagrams for the $qg \to qg$ channel. Mirror diagrams of (3), (5)
%  and (6) not shown. Figure from Ref.~\protect\cite{Kotko:2015ura}.}
%  \label{fig:qg2qg-diag}
%\end{minipage}%
%\begin{minipage}{.5\textwidth}
%  \centering
%\includegraphics[width=0.95\textwidth]{gg2qq-diag.png}
%%  \end{center}
%  \caption{Diagrams for the $gg \to q\qbar$ channel. Mirror diagrams of (3),
%  (5) and (6) not shown. Figure from Ref.~\protect\cite{Kotko:2015ura}.}
%  \label{fig:gg2qqbar-diag}
%\end{minipage}    
%\end{figure}

\begin{figure}%
    \centering
    \subfloat[$qg \to qg$]{{\includegraphics[width=0.47\textwidth]{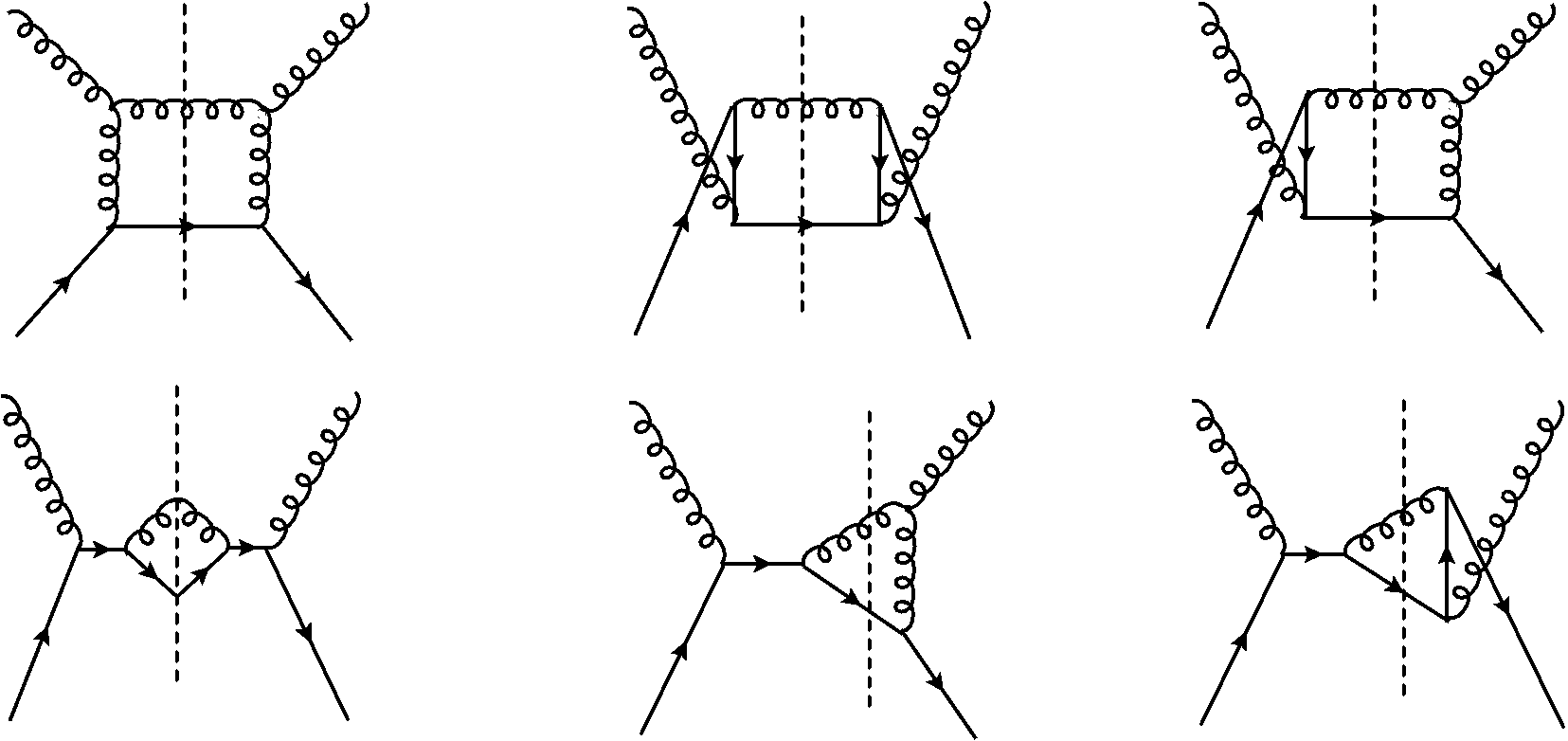}}}%
    \qquad
    \subfloat[$gg \to q\qbar$]{{\includegraphics[width=0.47\textwidth]{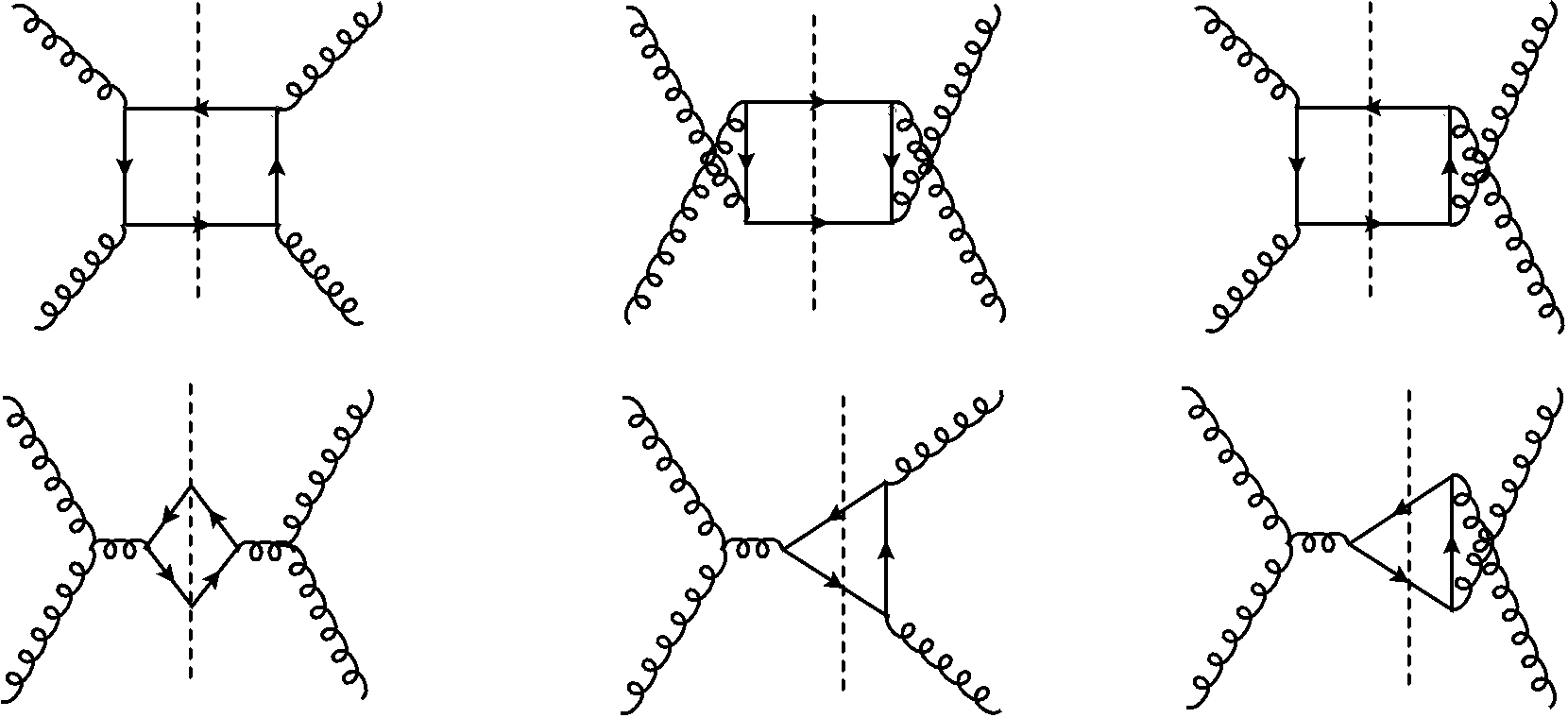}}}%
    \caption{Diagrams for the $qg \to qg$ and $gg \to q\qbar$ channels. Mirror diagrams are not shown. Figure from Ref.~\protect\cite{Kotko:2015ura}.}%
    \label{fig:qg}%
\end{figure}

\begin{figure}%
    \centering
    \subfloat{{\includegraphics[width=0.47\textwidth]{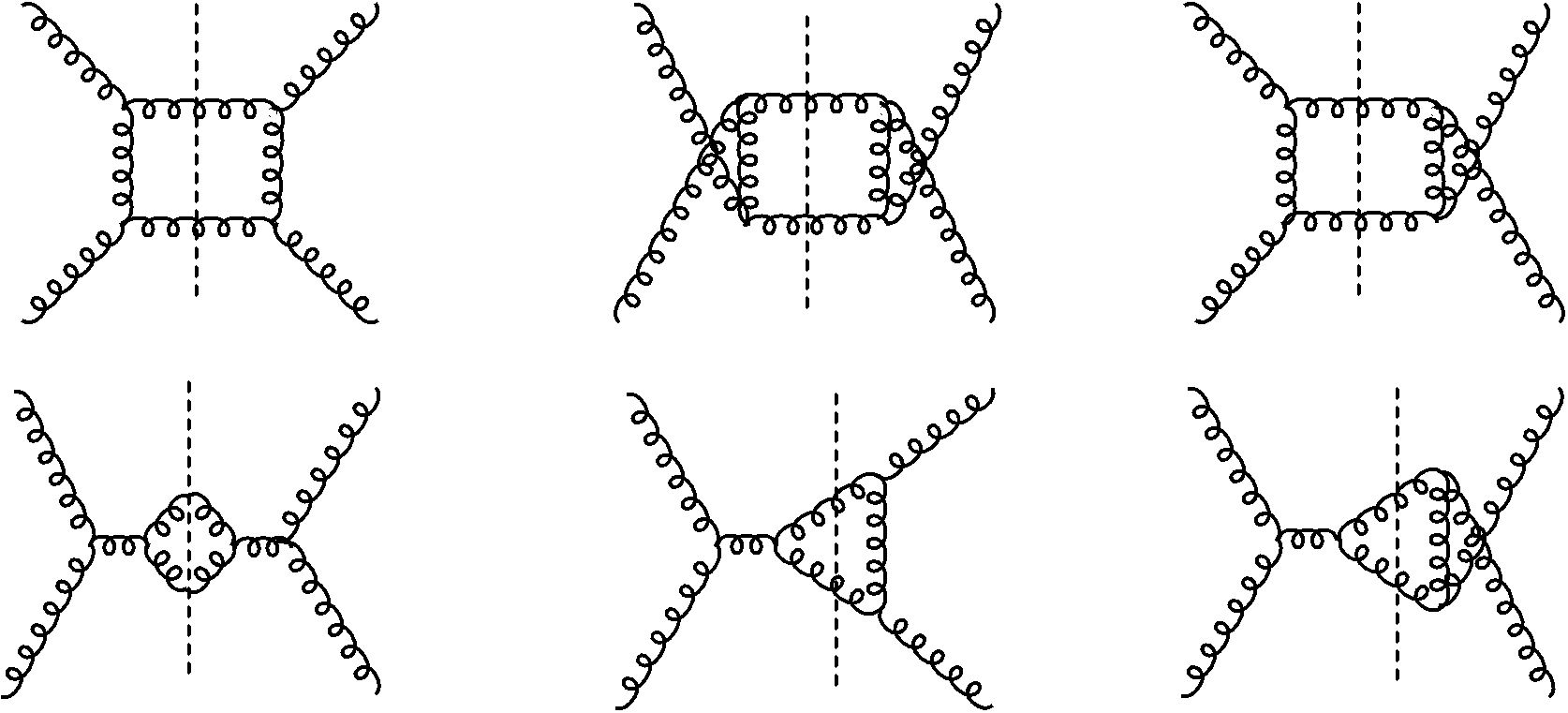}}}%
    \qquad
    \subfloat{{\includegraphics[width=0.47\textwidth]{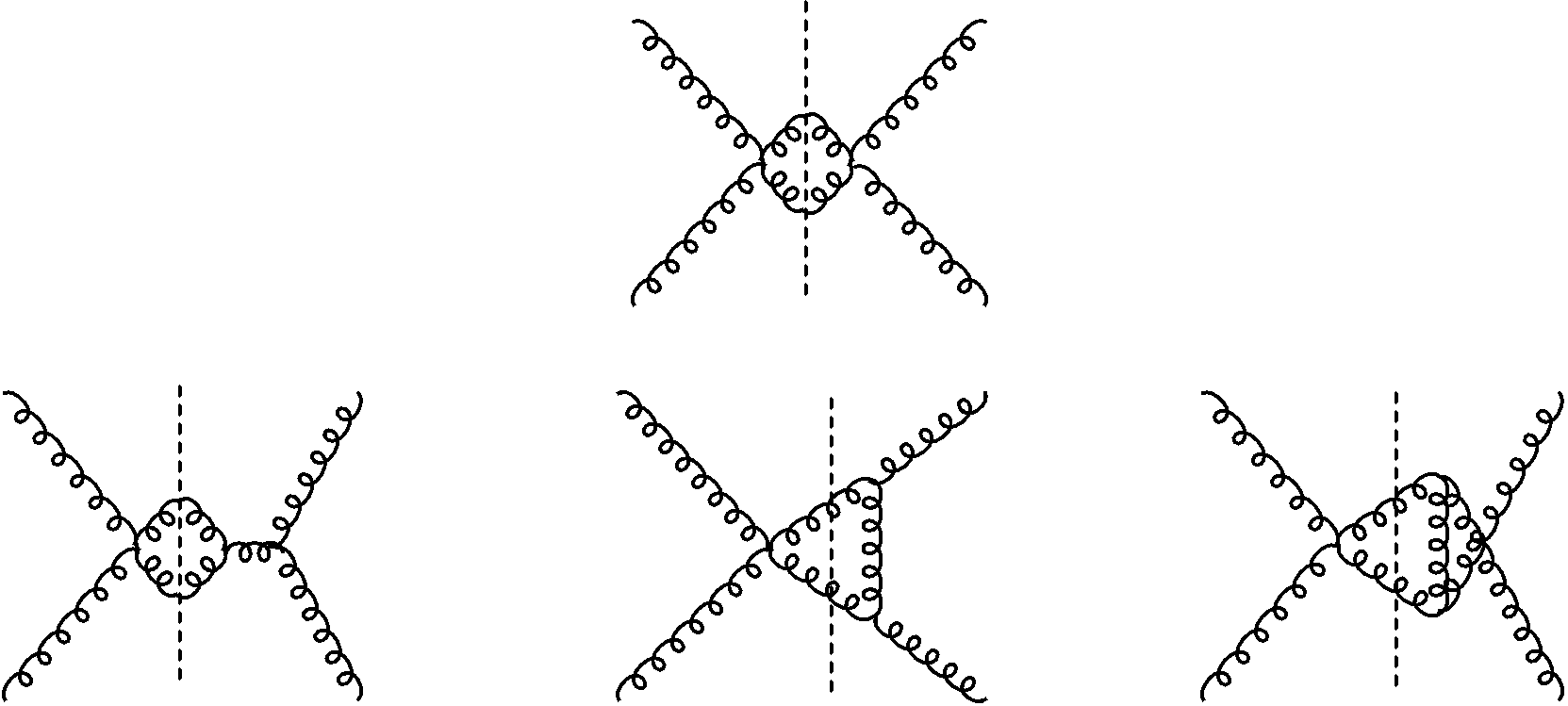}}}%
    \caption{Diagrams for the $gg \to gg$ channel with 3-gluon and 4-gluon
 vertex contributions. Mirror diagrams are not shown. Figure from Ref.~\protect\cite{Kotko:2015ura}.}%
    \label{fig:gg}%
\end{figure}

%\begin{figure}[t]
%  \begin{center}
%    \includegraphics[width=0.6\textwidth]{gg2qq-diag.png}
%  \end{center}
%  \caption{Diagrams for the $gg \to q\qbar$ channel. Mirror diagrams of (3),
%  (5) and (6) not shown. Figure from Ref.~\protect\cite{Kotko:2015ura}.}
%  \label{fig:gg2qqbar-diag}
%\end{figure}

%\begin{figure}[t]
%  \begin{center}
%    \includegraphics[width=0.6\textwidth]{gg2gg-3gdiag.png}
%  \end{center}
%  \caption{Diagrams for the $gg \to gg$ channel with only 3-gluon
%  vertices. Mirror diagrams of (3), (5) and (6) not shown. Figure from Ref.~\protect\cite{Kotko:2015ura}.}
%  \label{fig:gg2gg-3gdiag}
%\end{figure}
%
%\begin{figure}[t]
%  \begin{center}
%    \includegraphics[width=0.6\textwidth]{gg2gg-4gdiag.png}
%  \end{center}
%  \caption{Diagrams for the $gg \to gg$ channel with 4-gluon
%  vertex contributions. Mirror diagrams of (8), (9) and (10) not shown. Figure from Ref.~\protect\cite{Kotko:2015ura}.}
%  \label{fig:gg2gg-4gdiag}
%\end{figure}

The calculations of the perturbative partonic scattering and the non-perturbative TMD gluon distributions of the nucleus are related as the color structure of the $2\to2$ subprocess determines the type of TMD distribution associated with it. The $2\to2$ diagrams for the channels we are considering here ($qg \to qg$, $gg\to q\bar{q}$ and $gg\to gg$) are given in Figs.~\ref{fig:qg} and~\ref{fig:gg}. For each of these diagrams there is a definite TMD distribution related to it~\cite{Bomhof:2006dp}.

The unpolarized gluon TMDs are defined as Fourier transforms of forward matrix elements of bilocal products of the gluon field strength tensor~\cite{Collins:2002kn,12,Ji:2002aa,8}:
\begin{equation}
\mathcal{F}(x_2,k_t) =
2\int \frac{d\xi^+d^2{\boldsymbol\xi}}{(2\pi )^{3}p_A^{-}}e^{ix_2p_A^{-}\xi ^{+}-ik_t\cdot{\boldsymbol\xi}}
\left\langle p_A|\text{Tr}\left[F^{i-}\left( 0\right){\mathcal{U}}^{[\mathcal{C}]}_{\left[0,\xi\right]}F^{i-}\left(\xi\right){\mathcal{U}}^{[\mathcal{C'}]}_{\left[\xi,0\right]}
\right]|p_A\right\rangle,
\label{eq:tmd-UU'}
\end{equation}
with $x_2$ the longitudinal-momentum fraction and $k_t$ the transverse momentum of the gluon. The gauge links ${\mathcal{U}}^{[\mathcal{C}]}_{\left[0,\xi\right]}$ and ${\mathcal{U}}^{[\mathcal{C'}]}_{\left[\xi,0\right]}$ connecting the points $0$ and $\xi$ ensure a gauge invariant definition of the TMD distributions. They are path-ordered exponentials connecting the field strength tensors along a definite integration path $\mathcal{C}$ that depends on the partonic subprocess:
\be 
\mathcal{U}^{[\mathcal{C}]}_{\left[0,\xi\right]} = \mathcal{P} \exp\left[ -ig \int_\mathcal{C} dz \cdot A(z) \right] \, .
\ee
The gauge links are constituted by Wilson lines along the light-like plus direction (given with the light-like vector $n$) and along the transverse directions:
\bea 
&& U^n_{\left[a^+,b^+;\bf a
\right]}= \mathcal{P} \exp\left[ -ig \int_{a^+}^{b^+} dz^+\, A^-(z^+,z^-=0, \bf z= \bf a) \right]\, , \nonumber \\
&& U^T_{\left[a^+;\bf a,\bf b
\right]}=  \mathcal{P} \exp\left[ -ig \int_{\bf a}^{\bf b} d{\bf z} \cdot {\bf A}(z^+=a^+,z^-=0,\bf z) \right]\, .
\eea
One can define future pointing ($\mathcal{U}^{[+]}$) and past pointing ($\mathcal{U}^{[-]}$) gauge links:
\bea 
\mathcal{U}^{\left[ \pm\right] }_{\left[0,\xi\right]}= U^n_{\left[0^+,\pm\infty^+;\bf 0
\right]}
U^T_{\left[\pm\infty^+;\bf 0,{\boldsymbol{\xi}}
\right]}
U^n_{\left[\pm\infty^+,\xi^+;{\boldsymbol{\xi}}\right]}
\eea
and Wilson loops that are a product of $\mathcal{U}^{\left[ +\right] }$ and $\mathcal{U}^{\left[ -\right] }$:
\be
\mathcal{U}^{\left[\square \right] }=\mathcal{U}^{\left[ +\right] }_{\left[0,\xi\right]}\mathcal{U}^{\left[ -\right]\dagger}_{\left[0,\xi\right]}=\mathcal{U}^{\left[ -\right] \dagger}_{\left[0,\xi\right]}\mathcal{U}^{\left[ +\right]}_{\left[0,\xi\right]} \, .
\ee

\begin{figure}%
    \centering
    \subfloat{{\includegraphics[width=0.20\textwidth]{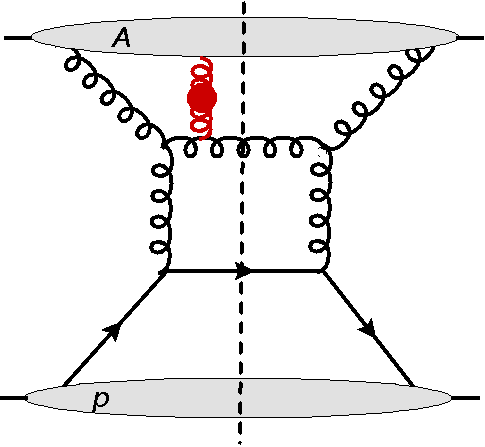}}}%
    \qquad
    \subfloat{{\includegraphics[width=0.20\textwidth]{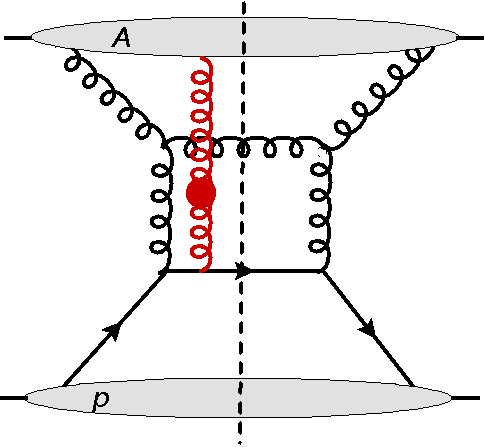}}}%
    \qquad
    \subfloat{{\includegraphics[width=0.20\textwidth]{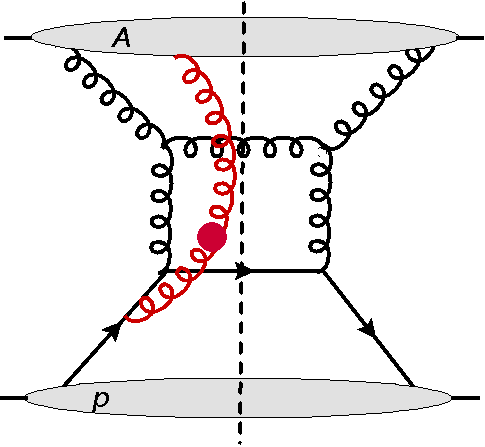}}}%
    \caption{Example of a $2\to2$ diagram with an arbitrary number of collinear gluons (in red) that need to be resummed into gauge links.}%
    \label{fig:GenericGaugeLink}%
\end{figure}

For the process we are considering here (dijet production in pA collisions) the path of integration of the gauge links is determined by the color flow in the $2\to 2 $ diagrams (the amplitude squared of the partonic subprocesses) shown in Figs.~\ref{fig:qg} and~\ref{fig:gg}. An example of a $2 \to2$ diagram is given in Fig.~\ref{fig:GenericGaugeLink}. We will call the gluon from the nucleus that participates in the interaction an active gluon (the gluon in the upper part of the diagrams in Figs.~\ref{fig:qg} and~\ref{fig:gg} and the black gluon in Fig.~\ref{fig:GenericGaugeLink}). For a gauge invariant definition of the TMD distributions, one needs to resum an arbitrary number of gluons from the nucleus that are collinear to the active gluon and that attach to the hard part (to the three partonic legs of the hard part except to the active one). These are the red gluons in Fig.~\ref{fig:GenericGaugeLink} where the circle denotes that there are an arbitrary number of collinear gluons attaching to the hard part (not just one). The resummation of the collinear gluons results in a Wilson line for each partonic leg. These Wilson lines get entangled with the amplitude for the $2\to2$ partonic scattering and there is no clear separation between the short-distance subprocess and the long-distance non-perturbative distributions. In order for the $2\to2$ amplitude to be isolated, the Wilson lines need to be absorbed in the definition of the matrix element of the TMD distributions in the nucleus. The process of disentangling of the Wilson lines from the hard part is affected by the color structure of the particular $2\to2$ diagram and different diagrams will result in different Wilson lines from which the gauge links ${\mathcal{U}}^{[\mathcal{C}]}_{\left[0,\xi\right]}$ and ${\mathcal{U}}^{[\mathcal{C'}]}_{\left[\xi,0\right]}$ are formed. The paths of integration $C$ and $C'$ are uniquely determined by the color structure of the subprocess and this is the origin of the process dependence of TMDs~\cite{Bomhof:2006dp}. When the disentangling of the Wilson lines from the hard part is not possible, the full cross section involves factorization-breaking contributions.

We can now see why the cross section for dijet production in pA collisions involves several TMD gluon distributions, including distributions that have been identified in other processes. The complicated color structure of the hard part (all of the partons in the initial and final state of the $2\to 2$ subprocess carry color and both initial and final state interactions are present) generates all the types of gauge links that we have defined above, $\mathcal{U}^{\left[ + \right] }$, $\mathcal{U}^{\left[ - \right] }$ and $\mathcal{U}^{\left[ \square \right] }$. For simpler processes, for example when there are (one or more) colorless particles in the initial or final state, or there are less partonic legs in the hard part, only one or a few of the mentioned types of gauge links will be relevant. 
%Studying dijet production in pA collisions then allows us to be thorough in the analysis of TMD gluon distributions at high energy as it contains all of the gauge links we have defined.

The gauge links for all types of $2\to2$ interactions, when the incoming and outgoing particles are either quarks or gluons, were calculated in Ref.~\cite{Bomhof:2006dp}. For the case when only incoming gluons are considered from the nucleus side, each diagram in Figs.~\ref{fig:qg} and~\ref{fig:gg} is associated with one of the following TMD distributions (different diagrams may give rise to the same distribution):
\begin{eqnarray}
\mathcal{F}_{qg}^{(1)}(x_2,k_t) &=&\frac{2}{p_A^-}\left[FT\right]_\xi
\left\langle p_A\left|\text{Tr}\left[ F^{i-}(\xi)\ \mathcal{U}^{\left[ -\right] \dagger }F^{i-}(0)\ \mathcal{U}^{\left[ +\right] }\right] \right|p_A\right\rangle\ , \nonumber
\label{eq:tmdqg1}\\
\mathcal{F}_{qg}^{(2)}(x_2,k_t) &=&\frac{2}{p_A^-}\left[FT\right]_\xi\
\frac{1}{N_c}\left\langle p_A\left|\text{Tr}\left[ F^{i-}(\xi)\ \mathcal{U}^{\left[ +\right] \dagger}F^{i-}(0)\ \mathcal{U}^{\left[ +\right] }\right]
\text{Tr}\left[ \mathcal{U}^{\left[\square \right] }\right]\right|p_A\right\rangle\ , \nonumber
\label{eq:tmdqg2}\\
\mathcal{F}_{gg}^{(1)}(x_2,k_t) &=&\frac{2}{p_A^-}\left[FT\right]_\xi\
\frac{1}{N_c}\left\langle p_A\left| \text{Tr}\left[ F^{i-}(\xi)\ \mathcal{U}^{\left[ -\right] \dagger }F^{i-}(0)\ \mathcal{U}^{\left[ +\right] }\right]
\text{Tr}\left[ \mathcal{U}^{\left[\square \right]\dagger }\right]\right|p_A\right\rangle\ , \nonumber
\\
\mathcal{F}_{gg}^{(2)}(x_2,k_t) &=&\frac{2}{p_A^-}\left[FT\right]_\xi\
\frac{1}{N_c}\left\langle p_A\left|\text{Tr}\left[ F^{i-}(\xi)\ \mathcal{U}^{\left[\square\right]\dagger} \right]
\textrm{Tr}\left[ F^{i-}(0)\ \mathcal{U}^{\left[ \square\right] }\right]\right|p_A\right\rangle\ , \nonumber
\\
\mathcal{F}_{gg}^{(3)}(x_2,k_t) &=&\frac{2}{p_A^-}\left[FT\right]_\xi
\left\langle p_A\left| \text{Tr}\left[F^{i-}(\xi)\ \mathcal{U}^{\left[+\right] \dagger }F^{i-}(0)\ \mathcal{U}^{\left[ +\right] }\right] \right|p_A\right\rangle\ , \nonumber
\\
\mathcal{F}_{gg}^{(4)}(x_2,k_t) &=& \frac{2}{p_A^-}\left[FT\right]_\xi
\left\langle p_A\left| \text{Tr}\left[F^{i-}(\xi)\ \mathcal{U}^{\left[-\right] \dagger }F^{i-}(0)\ \mathcal{U}^{\left[ -\right] }\right] \right|p_A\right\rangle\ , \nonumber
\\
\mathcal{F}_{gg}^{(5)}(x_2,k_t) &=& \frac{2}{p_A^-}\left[FT\right]_\xi
\left\langle p_A\left| \text{Tr}\left[F^{i-}(\xi)\ \mathcal{U}^{\left[\square \right] \dagger } \mathcal{U}^{\left[+\right] \dagger }
F^{i-}(0)\ \mathcal{U}^{\left[\square \right] } \mathcal{U}^{\left[ +\right] }\right] \right|p_A\right\rangle\ , \nonumber
\\
\mathcal{F}_{gg}^{(6)}(x_2,k_t) &=&\frac{2}{p_A^-}\left[FT\right]_\xi
\nonumber
 \\
&&{\hspace{-18pt}}\times \frac{1}{N^2_c}\left\langle p_A\left| \text{Tr}\left[ F^{i-}(\xi)\ \mathcal{U}^{\left[ +\right] \dagger }F^{i-}(0)\ \mathcal{U}^{\left[ +\right] }\right]
\text{Tr}\left[ \mathcal{U}^{\left[\square \right] }\right] \text{Tr}\left[ \mathcal{U}^{\left[\square \right] \dagger}\right] \right|p_A\right\rangle\ ,
\label{eq:tmdgg6}
\end{eqnarray}
where we used the notation $\left[FT\right]_\xi \equiv \int {d\xi^+d^2{\boldsymbol\xi}}/{(2\pi )^{3}}\,\,  e^{ix_2p_A^{-}\xi ^{+}-ik_t\cdot{\boldsymbol\xi}}$ for the Fourier transform, and $\mathcal{U}^{\left[ \pm \right] } \equiv \mathcal{U}^{\left[ \pm \right] }_{\left[0,\xi\right]}$. 

Each of these distributions $\mathcal{F}_{ag}^{(i)}$ is associated with a set of diagrams from Figs.~\ref{fig:qg} and~\ref{fig:gg}. We then define the hard factor $H_{ag\to cd}^{(i)}$, corresponding to a given $\mathcal{F}_{ag}^{(i)}$, as the sum of the amplitudes squared of all the diagrams associated with it. The hard factors are calculated with on-shell incoming partons (the dependence on $|k_t|$ survives only in the TMD gluon distributions). The on-shell hard factors were calculated in Ref.~\cite{Dominguez:2011wm} in the large-$N_c$ limit, and in Ref.~\cite{Kotko:2015ura} for a finite number of colors. We refer the reader to Ref.~\cite{Kotko:2015ura} for the explicit expressions.

We now have all the elements to write down an effective TMD factorization formula for the cross section for dijet production in pA collisions at forward rapidity in the correlation limit and for finite $N_c$~\cite{Kotko:2015ura}:
\begin{equation}
\frac{d\sigma^{pA\rightarrow {\rm dijets}+X}}{dy_1dy_2d^2p_{1t}d^2p_{2t}} =
\frac{\alpha _{s}^{2}}{(x_1x_2s)^{2}} \sum_{a,c,d} x_1 f_{a/p}(x_1) \sum_i H_{ag\to cd}^{(i)}\ \mathcal{F}_{ag}^{(i)}(x_2,k_t) 
\frac{1}{1+\delta_{cd}}\, .
\label{eq:tmd-main}
\end{equation}
In the next section we will show that the small-$x$ limit of the above formula coincides with the correlation limit of the CGC cross section.

With a similar analysis one can derive an effective TMD factorization for the other specific examples we are reviewing here:

\begin{enumerate}

%\subsection{Dijet production in DIS}

\item {\it Dijet production in DIS}. The TMD factorized cross section for dijet production in DIS, $\gamma^* + p/A\to q+\bar q+X$, in the correlation limit, involves only the $\mathcal{F}_{gg}^{(3)}$ gluon distribution~\cite{Dominguez:2011wm}:
\be 
\frac{d\sigma^{\gamma*+p/A\to q\bar qX}_{T/L}}{dy_1dy_2d^2p_{1t}d^2p_{2t}}
=
\delta\left(x_{\gamma*}-1\right)\mathcal{F}_{gg}^{(3)}(x_2,k_t) H^{\gamma*g\to q\bar q}_{T/L} \, ,
\label{eq:dijet_DIS}
\ee 
where $T/L$ stands for a transversely/longitudinally polarized photon. The hard factors $H^{\gamma*g\to q\bar q}_{T/L}$ can be found in Ref.~\cite{Dominguez:2011wm}. From the color flow of the subprocess it is easy to see why only the $\mathcal{F}_{gg}^{(3)}$ TMD is present in the cross section. Since only the final state carries color (the incoming photon is colorless) and there are only final state interactions between the quark-antiquark dipole and the proton/nucleus, the TMD involves only future pointing gauge links, $\mathcal{U}^{[+]}$. The distribution $\mathcal{F}_{gg}^{(3)}$ is the Weizs\"acker-Williams gluon distribution and it will be discussed in more detail in sections~\ref{sec:CGC} and~\ref{TMDsmall-x}. Dijet production in DIS with nuclei can be used for the extraction of the nuclear WW gluon TMD from future experimental facilities like the Electron Ion Collider (EIC) or the Large Hadron electron Collider (LHeC).

%\subsection{Photon-jet production in pp and pA collisions}

\item {\it Photon-jet production in pp and pA collisions.} The effective TMD factorization for photon-jet production, $q + p/A \to q + \gamma + X$, in pp or pA collisions involves only the $\mathcal{F}_{qg}^{(1)}$ gluon TMD:
\be 
\frac{d\sigma^{q+p/A\to \gamma qX}_{T/L}}{dy_1dy_2d^2p_{1t}d^2p_{2t}}
=
 \sum_{\text q~ flavor}x_1 f_{q/p}(x_1)\mathcal{F}_{qg}^{(1)}(x_2,k_t) H_{qg\to q\gamma}\, .
 \label{eq:TMD_gamma-jet}
\ee 
The hard factor $H_{qg\to q\gamma}$ can be found in Ref.~\cite{Dominguez:2011wm}. The initial state interactions between the incoming quark and the hadron or nucleus are resummed into a past pointing gauge link $\mathcal{U}^{[-]}$, while final state interactions of the outgoing quark are resummed into a future pointing gauge link $\mathcal{U}^{[+]}$~\cite{Bomhof:2006dp}, hence the appearance of $\mathcal{F}_{qg}^{(1)}$ in the cross section. The distribution $\mathcal{F}_{qg}^{(1)}$ is the dipole gluon distribution and it will be discussed in more detail in section~\ref{TMDsmall-x}. Besides photon-jet production, it also appears in DIS, SIDIS, Drell-Yan production, and photon, di-photon and hadron production in pp and pA collisions. 

%\subsection{Higgs boson production in pp and pA collisions}

\item {\it Higgs boson production in pp and pA collisions.} The cross section for Higgs boson production from gluon fusion, $g + p/A \to H + X$, in pp and pA collisions, in the hybrid approach, is~\cite{Sun:2011iw,Boer:2011kf,Schafer:2012yx}:
\be 
\frac{d\sigma^{g + p/A \to H + X}}{dyd^2p_{H}}
=
x_1 f_{g/p}(x_1)\mathcal{F}_{qg}^{(4)}(x_2,k_t)\sigma_0\, ,
\label{eq:TMD_Higgs}
\ee
where the hard part $\sigma_0$ is the leading-order cross section for scalar-particle production from two gluons. The TMD factorized cross section is valid for $p_{H} \ll M$, where $p_{H}$ and $M$ are the transverse momentum and the mass of the Higgs particle, respectively. In this process there are only initial state interactions of the incoming gluon with the proton or nucleus which are resummed in past pointing gauge links $\mathcal{U}^{[-]}$, hence the appearance of the $\mathcal{F}_{gg}^{(4)}$ TMD. The $\mathcal{F}_{gg}^{(4)}$ distribution will be discussed in more detail in section~\ref{TMDsmall-x}.

\end{enumerate}

\section{CGC Cross Section and Universality of the CGC Weight Functional}
\label{sec:CGC}

The CGC is an effective theory of QCD which describes the dynamics of partons in an ultra-relativistic proton or nucleus in the high gluon density regime~\cite{19,20,21,22}. When a proton or a nucleus is boosted to large momentum, its wave function is dominated by gluons with small values of $x$ and large enough lifetimes (because of time dilation) to be accessible by an external probe. If the energy of the nucleus is increased (if it is boosted to higher momentum) more gluons with smaller values of $x$ live long enough to be seen by the same external probe, {\it i.e.}\ as the energy increases, the density of gluons increases as well. In the CGC theory the fast degrees of freedom (partons with longitudinal momenta $k^- > \Lambda^-$ where $\Lambda^-$ is some momentum scale) are separated from the slow degrees of freedom (small-$x$ gluons with longitudinal momenta $k^- < \Lambda^-$). The change in the physics with a change in the separation scale $\Lambda^-$ is governed by the JIMWLK renormalization group equation~\cite{JIMWLK1,JIMWLK2,JIMWLK3,JIMWLK4,JIMWLK5,JIMWLK6}. The JIMWLK evolution equation is a non-linear equation which takes into account high-density effects that slow down the growth of the number of gluons with decreasing $x$. This saturation of gluon densities at high energy is characterized by the saturation momentum scale $Q_s$ which defines the transition between the linear and non-linear regimes. The saturation scale increases as $x$ decreases (as the hadron or nucleus is boosted to higher energy) and is larger for a nucleus than for a proton, $Q^2_s \sim A^{1/3}/x^{0.3}$, with $A$ the nuclear mass number. In collisions that involve an ultra-relativistic proton or nucleus, the saturation scale is larger than the QCD scale, $Q^2_s \gg \Lambda_{QCD}$, and determines the value of the running coupling constant, which is small in this regime, $\alpha_s(Q_s^2) \ll 1$. The smallness of the coupling constant allows for a perturbative calculation of observables.

The initial condition for the JIMWLK evolution at some starting scale $\Lambda_0^-$ is usually taken from the MV model~\cite{17,18}. The MV model is valid when quantum evolution effects are not large, but $x$ is sufficiently small such that the high-density regime is probed. Because of the high occupation number of color charges, the gluon field is treated as a classical field. The partons with $k^->\Lambda_0^-$ are considered as static sources for the classical gluon field $A^\mu$ with $k^-<\Lambda_0^-$. The field $A^\mu$ is calculated by solving the classical Yang-Mills equations in the presence of a current generated by the sources. The density of the sources per unit transverse area, $\rho$, is a stochastic variable that needs to be averaged among scattering events with a statistical distribution $\mathcal{W}[\rho]$. The distribution $\mathcal{W}[\rho]$ at some initial value of $\Lambda_0^-$ is the non-perturbative input in the problem, which is assumed to be Gaussian in the MV model. The solution of the Yang-Mills equations for a single ultra-relativistic proton or nucleus gives the classical Weizs\"acker-Williams gluon distribution, {\it i.e.}\ the number density of gluons in the hadronic or nuclear wave function.~\cite{JalilianMarian:1996xn,Kovchegov:1998bi} The WW distribution is peaked around the saturation scale, so most gluons have transverse momenta $k_t \approx Q_s$. When integrated over $k_t$, the number of gluons is finite, {\it i.e.}\ its rise is not infinite, but saturated.

The calculation of observables can be done first in the MV model, at the classical level, where one takes into account high-density effects and resums terms of the type $\alpha_s^2 A^{1/3}\sim 1$ while keeping $\Lambda^-=\Lambda^-_0$ fixed. Observables are averaged over the color field configurations with the distribution $\mathcal{W}[\rho]$. For some operator $\mathcal{O}[\rho]$ we have:
\be 
\langle \mathcal{O} \rangle \equiv \frac {\int \left[ D_\rho \right]  \mathcal{W}[\rho] \mathcal{O}[\rho]}{\int \left[ D_\rho \right]  \mathcal{W}[\rho]} \, .
\ee
Quantum corrections to the MV model are then introduced through the evolution of the distribution $\mathcal{W}_{\Lambda^-}[\rho]$ by changing the separation scale $\Lambda^-$ from $\Lambda_0^-$ to some new scale $\Lambda_1^-$ (the distribution now becomes dependent on $\Lambda^-$). This is done with the JIMWLK evolution equation which resums terms of the type $\alpha_s \ln \Lambda_0^-/\Lambda_1^- \sim 1$. The evolution equation can be written as:
\be 
\frac{\partial \mathcal{W}_{\Lambda^-}[\rho]}{\partial \ln \left ( \Lambda^- \right)}
=
H_{JIMWLK} \mathcal{W}_{\Lambda^-}[\rho]\, ,
\ee
where $H_{JIMWLK}$ is the JIMWLK Hamiltonian~\cite{JIMWLK1,JIMWLK2,JIMWLK3,JIMWLK4,JIMWLK5,JIMWLK6}.
% Repeating the evolution with a new scale, one can resum quantum corrections down to the scale $k^+=x_2 p_A^+$.

The CGC theory is universal in the sense that it can be applied to both hadrons and nuclei when the density of color charges in the wave function is sufficiently large. A proton with $A=1$ can be considered a dense system of gluons where saturation effects are important when it is boosted to large enough energy (or small enough $x$). For a nucleus, the onset of saturation effects happens faster as the saturation scale squared is enhanced by $A^{1/3}$.

Another universality property of the CGC is that the wave functional $\mathcal{W}_{\Lambda^-}[\rho]$ and its evolution with energy are independent of the process considered. The same form of $\mathcal{W}_{\Lambda^-}[\rho]$, with the same types of logarithms of energy resummed in its evolution, describes an ultra-relativistic proton or nucleus before the collision in any process, {\it i.e.}\ in DIS, pp, pA or AA collisions. Information about the CGC wave function extracted from one process can be applied to another. We will show below how this universality can be transferred to gluon TMDs at small $x$.

\subsection{Forward dijet production in pA collisions}

In this subsection we outline the main points of the calculation of the CGC cross section for dijet production in pA collisions. The kinematics that was described in subsection~\ref{sub:Kinematics} applies here as well with two differences. First, the CGC theory can be applied for any value of the momentum imbalance $k_t$ of the produced jets, so we consider any $k_t$ between $Q_s$ and $p_{1t}$, $p_{2t}$, and second, we assume that the values of $x_2$ that are probed in the nucleus are small enough such that we can apply the CGC theory. We again assume that the proton is dilute and described by PDFs, $f_{a/p}(x_1)$, in the hybrid approach.

\begin{figure}%
    \centering
    \subfloat[$q\to qg$  \label{fig:qqg}]{{\includegraphics[width=0.35\textwidth]{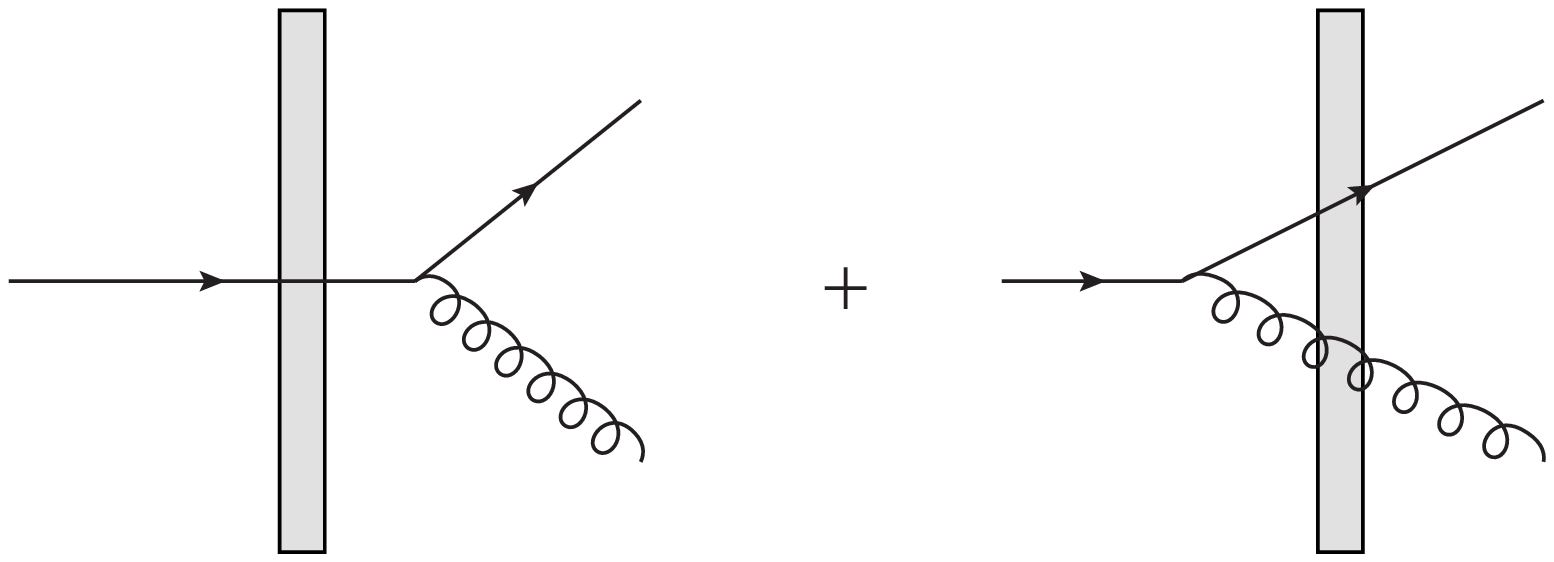}}}%
      \hspace{1cm}
    \qquad
    \subfloat[$g\to q\bar{q}$]{{\includegraphics[width=0.35\textwidth]{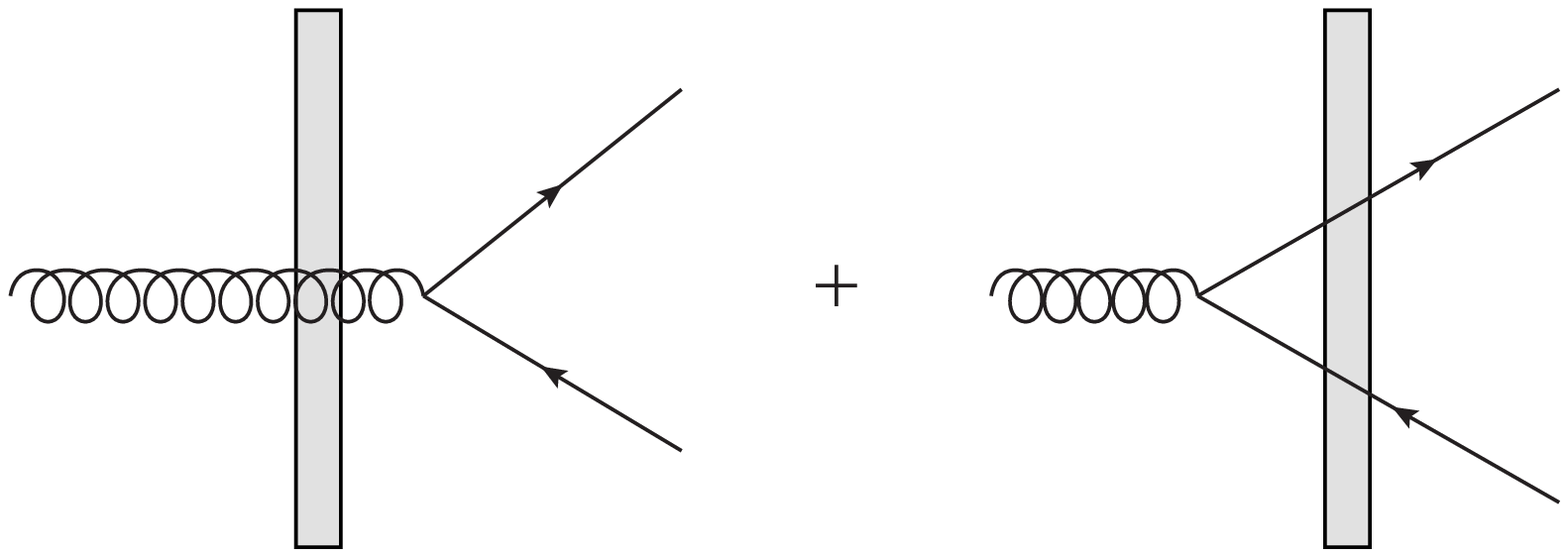}}}%
    \qquad
    \subfloat[$g\to gg$]{{\includegraphics[width=0.35\textwidth]{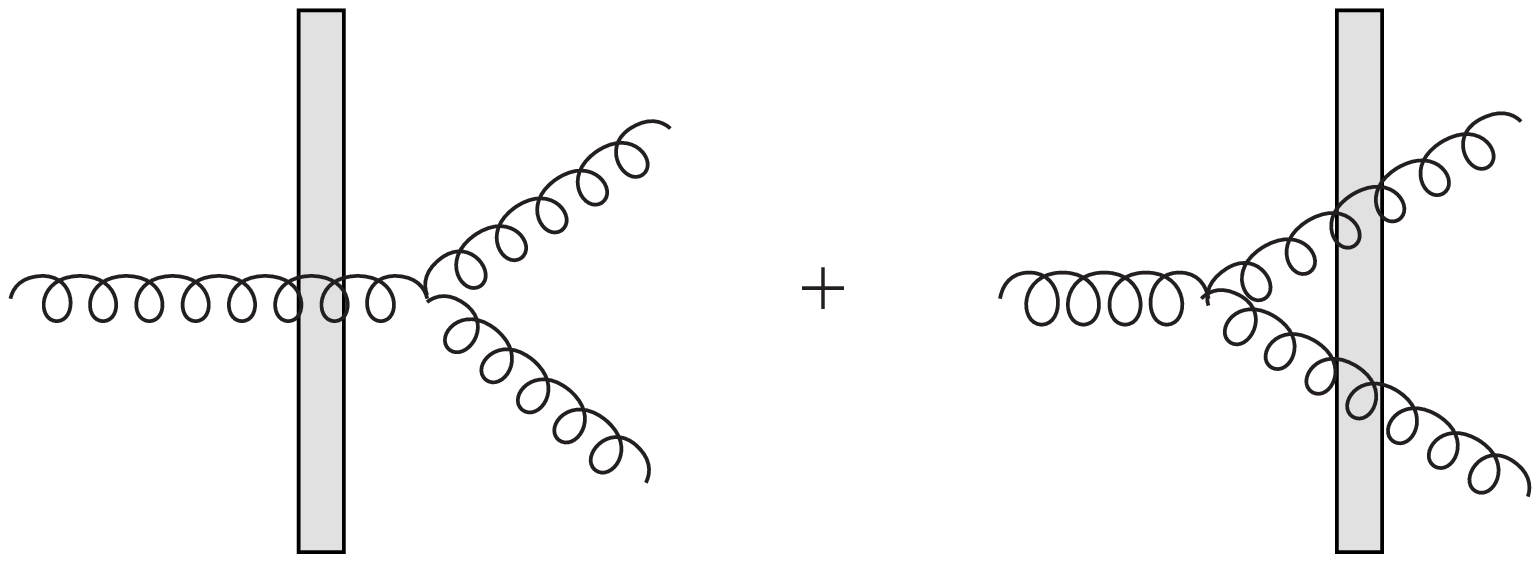}}}%
    \caption{Amplitudes for the three splitting channels, $q\to qg$, $g\to q\bar q$ and $g\to gg$, for dijet production in the CGC theory. The vertical bar represents the Lorentz contracted nucleus (the shockwave).}%
    \label{fig:ThreeChannelsCGC}%
\end{figure}

An incoming parton (a quark or a gluon) from the proton scatters off the dense gluon field in the nucleus. In the eikonal approximation, and in a covariant gauge, the multiple rescatterings of the incoming parton off the gluons can be resummed into a Wilson line running on the light cone from minus infinity to plus infinity. For an incoming quark the scattering is represented with a fundamental Wilson line, $U$, while for an incoming gluon with an adjoint Wilson line, $V$:
\bea
&& U_{\bf x}\equiv U^n_{[-\infty,+\infty;{\bf x}]}=\mathcal{P} \exp \left[- ig \int_{-\infty}^{\infty} dx^+ A_a^-(x^+, {\bf{x}}) t^a \right]\, ,
\nonumber \\
&& V_{\bf x}\equiv V^n_{[-\infty,+\infty;{\bf x}]}=\mathcal{P} \exp \left[- ig \int_{-\infty}^{\infty} dx^+ A_a^-(x^+, {\bf{x}}) T^a \right]\, ,
\label{eq:WilsonLines}
\eea
where $t^a$ and $T^a$ are the generators of the fundamental and adjoint representation of $SU(N)$, respectively.

The two jets in the final state are produced by the splitting of the incoming parton into two outgoing particles. We are working in a frame where the nucleus is ultra-relativistic and Lorentz contracted (sometimes referred to as a shockwave) and we consider splittings that happen either before or after the scattering off the nucleus, as splittings in the nucleus are suppressed by at least one power of the inverse center of mass energy~\cite{Blaizot:2004wu}. We consider three types of splittings, $q \to qg$, $g\to q\bar{q}$ and $g\to gg$, which correspond to the $2\to2$ scattering diagrams in the TMD factorization case. The amplitudes are schematically given in Fig.~\ref{fig:ThreeChannelsCGC}. The cross section is a convolution of light cone wave functions, describing the splitting of one parton into two, and correlators of Wilson lines, describing the scattering of the partons off the shockwave. The cross section for the $q\to qg$ case shown in Fig.~\ref{fig:qqg} in the $A^+=0$ gauge is~\cite{Marquet:2007vb}:
\bea
&&\frac{d\sigma(pA\to qgX)}{d^2p_{1t} d^2p_{2t}dy_1 dy_2} = \alpha_s C_F \left(1-z\right) p_1^+ x_1 f_{q/p}(x_1) \nonumber \\
&&\times
\int\frac{d^2{\bf x}}{(2\pi)^2}\frac{d^2{\bf x'}}{(2\pi)^2} 
\frac{d^2{\bf y}}{(2\pi)^2}\frac{d^2{\bf y'}}{(2\pi)^2}
e^{i p_{1t} \cdot ({\bf x'}-{\bf x})}
 e^{i p_{2t} \cdot ({\bf y'}-{\bf y})}
\sum_{\lambda\alpha\beta} \phi^{\lambda^*}_{\alpha\beta}(p,p_1^+,{\bf u'}) \phi^{\lambda}_{\alpha\beta}(p,p_1^+,{\bf u})
\nonumber \\
&&\times
\left\{S^{(4)}_{qg\bar{q}g}\left({\bf y},{\bf x},{\bf y'},{\bf x'}\right)-S^{(3)}_{qg\bar{q}}\left({\bf y},{\bf x},{\bf v'}\right)
-S^{(3)}_{qg\bar{q}}\left({\bf v},{\bf x'},{\bf y'}\right)+S^{(2)}_{q\bar{q}}\left({\bf v},{\bf v'}\right)\right\}\ ,
\label{eq:cgc-qg}
\eea
where $C_F$ is the Casimir of the fundamental representation. The transverse coordinates of the outgoing gluon (quark) in the amplitude and the complex conjugate amplitude are $\bf x$ and $\bf x'$ ($\bf y$ and $\bf y'$) respectively. We used the notation $\bf u=\bf x - \bf y$ for the transverse size of the quark-gluon dipole in the amplitude and ${\bf v} = z {\bf x} + (1-z) \bf y$ for the transverse position of the incoming quark in the amplitude, where $z=p_1^+/p^+$ is the longitudinal momentum fraction of the outgoing gluon with respect to the incoming quark. Similar notations hold for the complex conjugate amplitude. The function $\phi^{\lambda}_{\alpha\beta}(p,p_1^+,{\bf u})$ is the light cone wave function for a quark (with spin $\alpha$) splitting into a quark (with spin $\beta$) and a gluon (with polarization $\lambda$), which can be calculated in light cone perturbation theory. 
%We refer the reader to Ref.~\cite{cite} for the explicit expressions of the splitting functions for all the channels. 

The correlators $S^{(i)}$ describe the scattering of the partons off the shockwave. When the splitting $q\to qg$ happens before the scattering off the nucleus in both amplitude ($\mathcal M$) and complex conjugate amplitude ($\mathcal M^*$), the correlator $S^{(4)}_{qg\bar{q}g}\left({\bf y},{\bf x},{\bf y'},{\bf x'}\right)$ is a product of two fundamental and two adjoint Wilson lines describing the scattering of the outgoing quark and gluon off the shockwave in $\mathcal M$ and $\mathcal M^*$. When the incoming quark first scatters off the shockwave and then splits into a $qg$ pair, the correlator $S^{(2)}_{q\bar{q}}\left({\bf v},{\bf v'}\right)$ is a product of two fundamental Wilson lines for the scattering of the incoming quark in $\mathcal M$ and $\mathcal M^*$.
%(one for $\mathcal M$ and one for $\mathcal M^*$) for the outgoing quark and gluon respectively ($S^{(4)}_{qg\bar{q}g}\left({\bf y},{\bf x},{\bf y'},{\bf x'}\right))$. When the incoming quark first splits into a quark and a gluon and then scatters off the nucleus in both amplitude and complex conjugate amplitude, the correlator is a product of two fundamental Wilson lines for the incoming quark (one for $\mathcal M$ and one for $\mathcal M^*$), $S^{(2)}_{q\bar{q}}\left({\bf v},{\bf v'}\right)$. 
The interference terms involve two fundamental and one adjoint Wilson lines, $S^{(3)}_{qg\bar{q}}\left({\bf y},{\bf x},{\bf v'}\right)$ and $ S^{(3)}_{qg\bar{q}}\left({\bf v},{\bf x'},{\bf y'}\right)$. Their explicit expressions are:
\bea
S^{(4)}_{qg\bar{q}g}({\bf{y}},{\bf{x}},{\bf{y'}},{\bf{x'}})&=&\frac{1}{C_F N_c}\left<\text {Tr}\left(U_{\bf y}U^\dagger_{\bf y'}t^dt^c\right)
\left[V_{\bf x}V^\dagger_{\bf x'}\right]^{cd}\right>\ ,
\nonumber \\
S^{(3)}_{qg\bar{q}}({\bf{y}},{\bf{x}},{\bf{v'}})&=&\frac{1}{C_F N_c}\left<\text {Tr}\left(U^\dagger_{\bf v'}t^cU_{\bf y}t^d\right)V^{cd}_{\bf x}\right>\ ,
\nonumber \\
S^{(2)}_{q\bar{q}}({\bf{v}},{\bf{v'}})&=&\frac{1}{N_c}\left<\text {Tr} \left(U_{\bf v}U^\dagger_{\bf v'}\right)\right>\ .
\label{eq:corr}
\eea
The averages in the above correlators are performed over all possible configurations of color sources in the CGC wave function of the nucleus at a given value of $x_2$. They can be calculated in a model, for example with a Gaussian distribution $\mathcal{W}[\rho] \sim \exp[-\rho^2/\mu^2]$ as in the MV model, at some initial $x_0$. As a final step, their evolution towards small $x$ can be studied with the JIMWLK evolution equations. We emphasize that the cross section involves correlators of Wilson lines at four different transverse coordinates. Similar cross sections can be derived for the other two types of splittings $g\to q\bar q$ and $g \to gg$~\cite{Dominguez:2011wm,Iancu:2013dta}.

With a similar analysis one can derive CGC cross sections for the other specific examples we are reviewing here:

%\subsection{Dijet production in DIS}
\begin{enumerate}
\item {\it Dijet production in DIS.} The cross section for dijet production in DIS in the CGC theory is~\cite{Dominguez:2011wm}:
\bea
&&\frac{d\sigma^{\gamma*+p/A\to q\bar qX}_{T/L}}{d^3p_1 d^3p_2} = \alpha_{em} N_c e_q^2 \delta\left(p^+ - p_1^+ - p_2^+\right) \nonumber \\
&&\times
\int\frac{d^2{\bf x}}{(2\pi)^2}\frac{d^2{\bf x'}}{(2\pi)^2} 
\frac{d^2{\bf y}}{(2\pi)^2}\frac{d^2{\bf y'}}{(2\pi)^2}
e^{i p_{1t} \cdot ({\bf x'}-{\bf x})}
 e^{i p_{2t} \cdot ({\bf y'}-{\bf y})}
\sum_{\lambda\alpha\beta} \phi^{\lambda^*}_{\alpha\beta}(p,p_1^+,{\bf u'}) \phi^{\lambda}_{\alpha\beta}(p,p_1^+,{\bf u})
\nonumber \\
&&\times
\left\{S^{(4)}_{qg\bar{q}g}\left({\bf x},{\bf y},{\bf x'},{\bf y'}\right)
-S^{(2)}_{q\bar{q}}\left({\bf x},{\bf y}\right)-S^{(2)}_{q\bar{q}}\left({\bf x'},{\bf y'}\right)+ 1\right\}\ ,
\label{eq:cgc-DIS}
\eea
where $e_q$ is the electrical charge of the quark. The function $\phi^{\lambda}_{\alpha\beta}(p,p_1^+,{\bf u})$ is now the light cone wave function for a photon splitting into a quark-antiquark pair. The transverse coordinates have the same meaning as in the cross section for dijet production in pA collisions. The correlator $S^{(4)}_{qg\bar{q}g}\left({\bf x},{\bf y},{\bf x'},{\bf y'}\right)$ describes the scattering of the $q\bar q$ dipole in the amplitude and complex conjugate amplitude. The term equal to one in the curly brackets describes the passing of the photon through the nucleus with no interactions in $\mathcal{M}$ and $\mathcal{M^*}$. The interference terms are given by the $S^{(2)}$ correlators. 

%\subsection{Photon-jet production in pp and pA collisions}

\item {\it Photon-jet production in pp and pA collisions.} The cross section for direct photon-jet production in the CGC theory is\cite{Gelis:2002ki,Dominguez:2011wm}:
\bea
&&\frac{d\sigma^{q+p/A\to \gamma qX}_{T/L}}{d^3p_1 d^3p_2} = \alpha_{em} e_q^2 \delta\left(p^+ - p_1^+ - p_2^+\right) \nonumber \\
&&\times
\int\frac{d^2{\bf x}}{(2\pi)^2}\frac{d^2{\bf x'}}{(2\pi)^2} 
\frac{d^2{\bf y}}{(2\pi)^2}\frac{d^2{\bf y'}}{(2\pi)^2}
e^{i p_{1t} \cdot ({\bf x'}-{\bf x})}
 e^{i p_{2t} \cdot ({\bf y'}-{\bf y})}
\sum_{\lambda\alpha\beta} \phi^{\lambda^*}_{\alpha\beta}(p,p_1^+,{\bf u'}) \phi^{\lambda}_{\alpha\beta}(p,p_1^+,{\bf u})
\nonumber \\
&&\times
\left\{S^{(2)}_{q\bar{q}}\left({\bf y},{\bf y'}\right)
-S^{(2)}_{q\bar{q}}\left({\bf y},{\bf v'}\right)-S^{(2)}_{q\bar{q}}\left({\bf v},{\bf y'}\right)+ S^{(2)}_{q\bar{q}}\left({\bf v},{\bf v'}\right)\right\}\ .
\label{eq:cgc-photon-jet}
\eea
Now, the function $\phi^{\lambda}_{\alpha\beta}(p,p_1^+,{\bf u})$ is the light cone wave function for a $q\to q \gamma$ splitting. The transverse coordinates have the same meaning as in the cross section for dijet production in pA collisions. The correlator $S^{(2)}_{q\bar{q}}\left({\bf y},{\bf y'}\right)$ describes the scattering of the outgoing quark in the amplitude and complex conjugate amplitude, while the correlator $S^{(2)}_{q\bar{q}}\left({\bf v},{\bf v'}\right)$ describes the scattering of the incoming quark in the amplitude and complex conjugate amplitude. The other two terms, $S^{(2)}_{q\bar{q}}\left({\bf y},{\bf v'}\right)$ and $S^{(2)}_{q\bar{q}}\left({\bf v},{\bf y'}\right)$, are the interference terms.

\item {\it Higgs boson production in pp and pA collisions.} The cross section for Higgs boson production in pA collisions is~\cite{Mueller:2013wwa}:
\bea 
\frac{d\sigma^{g + p/A \to H + X}}{dyd^2p_{H}}
&=&
-\sigma_0 \, x_1 f_{g/p}(x_1) \nonumber \\
&~&\times \int\frac{d^2{\bf x}d^2{\bf x'}}{(2\pi)^2}
e^{i p_{H} \cdot ({\bf x'}-{\bf x})}
\left\langle \text{Tr}\left[ (\partial_i U_{\bf x}) U^\dagger_{{\bf x}'} (\partial_i U_{{\bf x}'}) U^\dagger_{\bf x}  \right] \right\rangle
\, ,
\label{eq:Higgs}
\eea
where $\sigma_0 = g_\Phi^2/\left(4 g^2(N_c^2-1)(1-\epsilon)\right) $ with $g_\Phi$ the effective Higgs coupling to the gluon field and $\epsilon$ dimensional regularization parameter.

\end{enumerate}

%\subsection{Higgs boson production in pp and pA collisions}
%
%The CGC cross section for Higgs production in the hybrid approach was calculated in Ref.~ref\cite{} and it was found that involves a correlator of the gluon field strength tensor with past pointing Wilson lines, {\it i.e.}\ it was concluded that the cross section is of the same form as the cross section in the TMD factorization framework.
%%
%\bea 
%\frac{d\sigma^{g + p/A \to H + X}_{T/L}}{dyd^2p_{H}}
%&=&
%-\sigma_0 x_1 f_{g/p}(x_1) \nonumber \\
%&\times &
%\int\frac{d^2{\bf x}}{(2\pi)^2}\frac{d^2{\bf x'}}{(2\pi)^2}
%e^{i p_{H} \cdot ({\bf x'}-{\bf x})}
%\left\langle \text{Tr}\left[ (\partial_i U_{\bf x}) U^\dagger_{\bf x'} (\partial_i U_{\bf x'}) U^\dagger_{\bf x}  \right] \right\rangle_{x_2} \, ,
%\eea

\section{Equivalence Between the TMD Factorization and the CGC Cross Section}
\label{sec:equiv}

In this section we will discuss the equivalence between the effective TMD factorization formulas from section~\ref{sec:TMD} and the CGC expressions from section~\ref{sec:CGC} in their overlapping domain of validity. As before, we consider dijet production in pA collisions in detail and then extend the conclusions to the other processes.

As discussed in the introduction, there exists a particular kinematic region for dijet production in pA collisions where the two independent frameworks of TMD factorization and the CGC effective theory are valid and should yield the same results. In the first subsection we will consider the small-$x$ limit of the TMD factorization formula~(\ref{eq:tmd-main}), while in the second subsection we will discuss the correlation limit of the CGC cross section~(\ref{eq:cgc-qg}). At the end, we will explain the equivalence between the two theories.

\subsection{Small-$x$ limit of gluon TMDs}

Taking the small-$x$ limit of the TMD cross section~(\ref{eq:tmd-main}) amounts to taking the small-$x$ limit of the TMD gluon distributions $\mathcal{F}_{ag}^{(i)}$. We start with the definition~(\ref{eq:tmd-UU'}), with the hadronic/nuclear state with momentum $p_A$ normalized as $\langle p|p'\rangle=(2\pi )^{3}\ 2p^-\delta(p^- - p'^-)\delta^{(2)}(p_t - p'_t)$, and we use translational invariance to write:
\bea
 &&\hspace{-20pt}\mathcal{F}(x_2,k_t)=
2\int \frac{d\xi^+d^2{\boldsymbol\xi}}{(2\pi )^{3}p_A^{-}}e^{ix_2p_A^{-}\xi ^{+}-ik_t\cdot{\boldsymbol\xi}}
\left\langle p_A|\text{Tr}\left[F^{i-}_0{\mathcal{U}}^{[\mathcal{C}]}_{\left[0,\xi\right]}F^{i-}_\xi{\mathcal{U}}^{[\mathcal{C'}]}_{\left[\xi,0\right]}
\right]|p_A\right\rangle \nonumber \\
&&
\hspace{-20pt}=\frac{4}{\langle p_A | p_A \rangle} \int \frac{d^3x d^3y}{(2\pi )^{3}}e^{ix_2p_A^{-}(x ^{+}-y^+)-ik_t\cdot({\boldsymbol x}-{\boldsymbol y})} 
\left\langle p_A|\text{Tr}\left[F^{i-}_x{\mathcal{U}}^{[\mathcal{C}]}_{\left[x,y\right]}F^{i-}_y{\mathcal{U}}^{[\mathcal{C'}]}_{\left[y,x\right]}
\right]|p_A\right\rangle \hspace{-3pt},
\eea
where we defined $F^{i-}_x \equiv F^{i-}(x)$. In the small-$x$ limit we replace the average over the state $p_A$ with an average over the CGC wave function of an ultra-relativistic proton or nucleus at the longitudinal-momentum fraction $x_A$, which is the smallest longitudinal-momentum fraction probed, determined by the kinematics of the process, {\it i.e.}\ $\langle p_A| \cdots | p_A \rangle / \langle p_A | p_A \rangle \to \langle \cdots \rangle_{x_A}$~\cite{Dominguez:2011wm}:
\be 
 \mathcal{F}(x_2,k_t) =
 4 \int \frac{d^3x d^3y}{(2\pi )^{3}}e^{ix_2p_A^{-}(x ^{+}-y^+)-ik_t\cdot({\boldsymbol x}-{\boldsymbol y})}
\left\langle \text{Tr}\left[F^{i-}_x{\mathcal{U}}^{[\mathcal{C}]}_{\left[x,y\right]}F^{i-}_y{\mathcal{U}}^{[\mathcal{C'}]}_{\left[y,x\right]}
\right]\right\rangle _{x_A} . 
\ee
The normalization factor used for this replacement will be evident below when it will lead to a complete agreement with the CGC cross section. When we consider the JIMWLK evolution of the TMDs in the next section, we will set $x_A = x_2$. In order to be able to perform the integration over $x^+$ and $y^+$ we  set $\exp[{ix_2p_A^{-}(x ^{+}-y^+)}]=1$ in the small-$x_2$ limit. This will also allow us to study the JIMWLK evolution of all the TMDs we have encountered before. Evolution equations that are valid for any value of $x$ have been derived in Ref.~\cite{Balitsky:2015qba} for $\mathcal{F}_{gg}^{(3)}$ and in Ref.~\cite{Balitsky:2016dgz} for $\mathcal{F}_{gg}^{(4)}$; it is not known how to solve them however. We comment on those equations later in the review. Here, we want to study and compare the JIMWLK evolution of all the TMDs, so we set $\exp[{ix_2p_A^{-}(x ^{+}-y^+)}]=1$ and rewrite the TMDs from Eq.~(\ref{eq:tmdgg6}) in the $A^+=0$ gauge as~\cite{Marquet:2016cgx}: 
\begin{eqnarray}
\mathcal{F}_{qg}^{(1)}(x_2,k_t) &=&\frac{4}{g^2}\int\frac{d^2{\bf x} d^2{\bf y}}{(2\pi)^3}\ e^{-ik_t\cdot({\bf x}-{\bf y})}
\left\langle \text{Tr}\left[ (\partial_i U^\dagger_{\bf x}) (\partial_i U_{\bf y})\right] \right\rangle_{x_2}
\label{eq:fqg1}\nonumber  \\
\mathcal{F}_{qg}^{(2)}(x_2,k_{t}) &=& -\frac{4}{g^2}\int\frac{d^2{\bf x} d^2{\bf y}}{(2\pi)^3}e^{-ik_t\cdot({\bf x}-{\bf y})}
\frac1{N_c}
\left\langle \text{Tr}\left[ (\partial_i U_{\bf x}) U^\dagger_{\bf y} (\partial_i U_{\bf y}) U^\dagger_{\bf x}  \right] \text{Tr}\left[ U_{\bf y} U^\dagger_{\bf x}\right] \right\rangle_{x_2}
\label{eq:fqg2}\nonumber \\
\mathcal{F}_{gg}^{(1)}(x_2,k_{t}) &=& \frac{4}{g^2}\int\frac{d^2{\bf x} d^2{\bf y}}{(2\pi)^3}\ e^{-ik_t\cdot({\bf x}-{\bf y})}
\frac1{N_c}
\left\langle \text{Tr}\left[  (\partial_i U_{\bf y}) (\partial_i U^\dagger_{\bf x}) \right] \text{Tr}\left[ U_{\bf x} U^\dagger_{\bf y}\right] \right\rangle_{x_2}
\label{eq:fgg1}\nonumber \\
\mathcal{F}_{gg}^{(2)}(x_2,k_{t}) &=& -\frac{4}{g^2}\int\frac{d^2{\bf x} d^2{\bf y}}{(2\pi)^3}\ e^{-ik_t\cdot({\bf x}-{\bf y})}
\frac1{N_c}
\left\langle \text{Tr}\left[ (\partial_i U_{\bf x}) U^\dagger_{\bf y}\right] \text{Tr}\left[ (\partial_i U_{\bf y}) U^\dagger_{\bf x}  \right] \right\rangle_{x_2}
\label{eq:fgg2}\nonumber \\
\mathcal{F}_{gg}^{(3)}(x_2,k_t) &=&
- \frac{4}{g^2}\int\frac{d^2{\bf x} d^2{\bf y}}{(2\pi)^3}\ e^{-ik_t\cdot({\bf x}-{\bf y})}
\left\langle \text{Tr}\left[ (\partial_i U_{\bf x}) U^\dagger_{\bf y} (\partial_i U_{\bf y}) U^\dagger_{\bf x}  \right] \right\rangle_{x_2} 
\label{eq:fgg3}\nonumber \\
\mathcal{F}_{gg}^{(4)}(x_2,k_{t}) &=&-\frac{4}{g^2}\int\frac{d^2{\bf x} d^2{\bf y}}{(2\pi)^3}\ e^{-ik_t\cdot({\bf x}-{\bf y})}
\left\langle \text{Tr}\left[ (\partial_i U_{\bf x}) U^\dagger_{\bf x} (\partial_i U_{\bf y}) U^\dagger_{\bf y}  \right] \right\rangle_{x_2}
\label{eq:fgg4}\nonumber \\
\mathcal{F}_{gg}^{(5)}(x_2,k_{t}) &=& -\frac{4}{g^2}\int\frac{d^2{\bf x} d^2{\bf y}}{(2\pi)^3}\ e^{-ik_t\cdot({\bf x}-{\bf y})}
\left\langle \text{Tr}\left[ (\partial_i U_{\bf x}) U^\dagger_{\bf y} U_{\bf x} U^\dagger_{\bf y} (\partial_i U_{\bf y}) U^\dagger_{\bf x} U_{\bf y} U^\dagger_{\bf x} \right] \right\rangle_{x_2}
\label{eq:fgg5}\nonumber \\
\mathcal{F}_{gg}^{(6)}(x_2,k_{t}) &=& -\frac{4}{g^2}\int\frac{d^2{\bf x} d^2{\bf y}}{(2\pi)^3}\ e^{-ik_t\cdot({\bf x}-{\bf y})} \nonumber \\
&&{\hspace{-10pt}}\times
\frac1{N_c^2}
\left\langle \text{Tr}\left[ (\partial_i U_{\bf x}) U^\dagger_{\bf y} (\partial_i U_{\bf y}) U^\dagger_{\bf x}  \right] \text{Tr}\left[ U_{\bf x} U^\dagger_{\bf y}\right] \text{Tr}\left[ U_{\bf y} U^\dagger_{\bf x}\right] \right\rangle_{x_2},
\label{eq:fgg6} 
\end{eqnarray}
where we used for the derivative of a Wilson line in the $A^+=0$ gauge:
\be 
\partial^i U_{\bf y} = -ig \int_{-\infty}^{\infty} dy^+ U^n_{[-\infty, y^+;{\bf y}]} F^{i-} (y) U^n_{[y^+, +\infty;{\bf y}]} \, .
\ee
The expressions for the TMDs are simplest in the $A^+=0$ gauge where transverse Wilson lines are unity. For a gauge invariant derivation of $\mathcal{F}_{qg}^{(1)}$ see {\it e.g.}\ Ref.~\cite{Boer:2016xqr}. Note that the fundamental Wilson lines $U_{\bf x}$ run from minus infinity to plus infinity, as given in Eq.~(\ref{eq:WilsonLines}).
%The distribution $\mathcal{F}_{qg}^{(1)}$ is the dipole gluon distribution (sometimes denoted as $xG^{(2)}$) and $\mathcal{F}_{gg}^{(3)}$ is the WW distribution (sometimes denoted as $xG^{(1)}$). We will discuss these two distributions in particular in section~\ref{TMDsmall-x}.
Using~(\ref{eq:fgg6}) in Eqs.~(\ref{eq:tmd-main}),~(\ref{eq:dijet_DIS}),~(\ref{eq:TMD_gamma-jet}) and~(\ref{eq:TMD_Higgs}) in section~\ref{sec:TMD} we obtain the small-$x$ limit of the cross sections for dijet production in pp and pA collisions and in DIS, photon-jet production in pp and pA collisions and Higgs production in pp and pA collisions. 

\subsection{Correlation limit of the CGC cross section}

We study the correlation limit of the CGC cross section for the $q \to qg$ channel (the same conclusions can be drawn from the other two cases). The large transverse momenta of the final jets can come from two mechanisms. First, the quark can split into a quark and a gluon with large transverse momenta already in the incoming wave function, such that the hard momenta of the final jets come from the splitting itself. Second, if the incoming quark splits into a quark and a gluon with small transverse momenta, they both need to receive large momentum transfer from the nucleus in order to produce two hard jets in the final state. It was shown in Ref.~\cite{Altinoluk:2011qy} that the first mechanism dominates the cross section and thus we focus the discussion on this case only. After the splitting, the quark and the gluon can both exchange soft momenta with the gluons in the nucleus and they will propagate almost back-to-back in the plane transverse to the collision axis. The transverse momentum imbalance of the jets in this case is small and is sensitive to the transverse momenta of the small-$x$ gluons in the nucleus. This corresponds to the correlation limit. (The case when only one of the outgoing particles scatters off the nucleus with a large momentum exchange will be analyzed in subsection~\ref{sub:ITMD}.) 

In this picture, the momentum of the jets is conjugate to the transverse size of the outgoing quark-gluon dipole. As the momentum of the jets is the hard scale in the process (with the other momentum scales being softer), the correlation limit corresponds to small dipole sizes compared to the other transverse distances in the problem. The correlation limit of the CGC cross section is then obtained by expanding Eq.~(\ref{eq:cgc-qg}) for small $|\bf u|$ and $|\bf u'|$. To leading order in $|\bf u|$ and $|\bf u'|$ the correlators in Eq.~(\ref{eq:corr}) become correlators of Wilson lines and first order derivatives of Wilson lines at only two transverse positions ($|\bf v|$ and $|\bf v'|$), resulting in the correlators that define the TMDs in Eq.~(\ref{eq:fgg6}). After performing the integrals over $|\bf u|$ and $|\bf u'|$ one can show that to leading order in $|\bf u|$ and $|\bf u'|$ the cross section in the CGC theory (Eq.~(\ref{eq:cgc-qg}) plus the cross sections for the other two channels) coincides completely with the TMD factorization formula (\ref{eq:tmd-main}) with the TMDs at small $x$ in Eq.~(\ref{eq:fgg6}). We conclude that the CGC theory contains the TMD factorization result as a leading power in the inverse hard scale and that gluon TMDs at small $x$ can be identified with UGDs in the CGC theory.

For the other processes discussed in section~\ref{sec:TMD} the analysis goes along the same lines. For the cases of dijet production in DIS and jet-photon production in pp and pA collisions, when the momentum imbalance of the final state particles is much smaller than their transverse momenta, the correlators in the CGC expressions simplify and, to leading order in the hard scale, the cross sections reduce to the corresponding TMD factorization formulas. For the case of Higgs boson production in pp and pA collisions, the CGC cross section, in the hybrid approach, was calculated in Ref.~\cite{Schafer:2012yx} and it was found that it involves a correlator of the gluon field strength tensor with past pointing Wilson lines, {\it i.e.}\ it was concluded that the cross section is the same as the TMD factorization result involving $\mathcal{F}_{gg}^{(4)}$. This is in agreement with the result~(\ref{eq:Higgs}) from Ref.~\cite{{Mueller:2013wwa}} when $\mathcal{F}_{gg}^{(3)} =\mathcal{F}_{gg}^{(4)}$ (see section~\ref{TMDsmall-x}). For similar analysis for SIDIS in electron-proton or electron-nucleus collisions see {\it e.g.}\ Ref.~\cite{Marquet:2009ca}.

The correlation limit of the CGC cross section can be derived for other processes as well, and the final result can be written in a factorized form with hard factors and gluon TMDs as CGC averages. For instance, Ref.~\cite{Akcakaya:2012si} derived the correlation limit of the CGC cross section for quark-antiquark pair production in hadron-hadron collisions and an agreement with the TMD factorization approach was found, Ref.~\cite{Benic:2017znu} derived the correlation limit for photon-jet production at central rapidites from the partonic subprocess $g \to q \bar q \gamma$ and Ref.~\cite{Altinoluk:2018uax} derived the correlation limit for dijet production in pA collisions accompanied with an extra soft photon in the final state. The TMDs that appear in these cases are linear combinations of the TMDs in Eq.~(\ref{eq:fgg6}). Because of the universality of the CGC theory, conclusions about gluon TMDs extracted from one process can be applied to other processes as well.

\section{WW and Dipole Gluon TMDs}
\label{TMDsmall-x}

The WW and dipole distributions are usually referred to as the two fundamental distributions in the CGC. One reason for this is that the WW distribution can be related to a bilocal product of the gauge field in the light cone gauge and is the proper number density of gluons in the hadronic or nuclear wave function, while the dipole distribution is the one that appears in most CGC cross sections. The second reason is that in the large $N_c$ limit the distributions $\mathcal{F}_{qg}^{(2)}$, $\mathcal{F}_{gg}^{(1)}$, $\mathcal{F}_{gg}^{(2)}$ and $\mathcal{F}_{gg}^{(6)}$ can be written as convolutions of the two fundamental ones~\cite{Dominguez:2011wm}. Finally, until now, only the WW and dipole distributions are known to appear individually in cross sections, such that they can be disentangled from the other TMDs and can be extracted from experiments.

The WW distribution (frequently called $xG^{(1)}$ or $f_1^{g[+,+]}$) in the TMD language is defined as:
\be 
\mathcal{F}_{gg}^{(3)}(x_2,k_t) =2\int \frac{d\xi^+d^2{\boldsymbol\xi}}{(2\pi )^{3}p_A^{-}}e^{ix_2p_A^{-}\xi ^{+}-ik_t\cdot{\boldsymbol\xi}}
\left\langle p_A\left| \text{Tr}\left[F^{i-}(\xi)\ \mathcal{U}^{\left[+\right] \dagger }F^{i-}(0)\ \mathcal{U}^{\left[ +\right] }\right] \right|p_A\right\rangle ,
\ee
while at small $x$, in a covariant gauge, it is:
\be 
\mathcal{F}_{gg}^{(3)}(x_2,k_t) =
- \frac{4}{g^2}\int\frac{d^2{\bf x} d^2{\bf y}}{(2\pi)^3}\ e^{-ik_t\cdot({\bf x}-{\bf y})}
\left\langle \text{Tr}\left[ (\partial_i U_{\bf x}) U^\dagger_{\bf y} (\partial_i U_{\bf y}) U^\dagger_{\bf x}  \right] \right\rangle_{x_2}\, .
\ee 
It behaves as $\sim Q_s^2/k_t^2$ at large $k_t$, $k_t \gg Q_s$, and as $\sim \ln Q_s^2/k_t^2$ at small $k_t$. It has been proposed that the WW gluon distribution can be observed in quark-antiquark jet correlations in DIS~\cite{Dominguez:2011wm}.

The other gluon distributions that we have defined, do not have a number density interpretation. They result from the requirement of a factorized form of the cross section and they are not parton distributions in the real sense. Nevertheless, they can provide valuable information on saturation effects in high-energy collisions. In fact, the dipole gluon distribution (frequently called $xG^{(2)}$ or $f_1^{g[+,-]}$) is the one that appears in most of the observables in high-energy scatterings. It is related to the $S$-matrix for a quark-antiquark dipole scattering off the CGC field of a proton or a nucleus in DIS and in the TMD language it is defined as:
\be 
\mathcal{F}_{qg}^{(1)}(x_2,k_t) =2\int \frac{d\xi^+d^2{\boldsymbol\xi}}{(2\pi )^{3}p_A^{-}}e^{ix_2p_A^{-}\xi ^{+}-ik_t\cdot{\boldsymbol\xi}}
\left\langle p_A\left|\text{Tr}\left[ F^{i-}(\xi)\ \mathcal{U}^{\left[ -\right] \dagger }F^{i-}(0)\ \mathcal{U}^{\left[ +\right] }\right] \right|p_A\right\rangle\ \, 
\ee
while at small $x$ it is related to a gauge invariant Wilson loop~\cite{Boer:2016xqr}:
\bea
\mathcal{F}_{qg}^{(1)}(x_2,k_t)& =&\frac{4}{g^2}\int\frac{d^2{\bf x} d^2{\bf y}}{(2\pi)^3}\ e^{-ik_t\cdot({\bf x}-{\bf y})}
\left\langle \partial_{\bf x}^i \partial_{\bf y}^i \text{Tr}\left[ {\mathcal{U}}^{[\square]}\right] \right\rangle_{x_2} \nonumber \\
& =&
\frac{4k_t^2}{g^2}\int\frac{d^2{\bf x} d^2{\bf y}}{(2\pi)^3}\ e^{-ik_t\cdot({\bf x}-{\bf y})}
\left\langle  \text{Tr}\left[ {\mathcal{U}}^{[\square]}\right] \right\rangle_{x_2} \, .
\eea
In a covariant gauge the transverse Wilson lines in the loop operator are unity and the dipole distribution reduces to the form in Eq.~(\ref{eq:fgg5}). It behaves the same as the WW distribution at large $k_t$, $\sim Q_s^2/k_t^2$, while at small $k_t$ it behaves as $\sim k_t^2$.

In the small-$x$ limit the TMD with two past-pointing gauge links, $\mathcal{F}_{gg}^{(4)}$, reduces to the WW distribution as well, however, in the most general case $\mathcal{F}_{gg}^{(3)}$ and $\mathcal{F}_{gg}^{(4)}$ are not the same and they obey different evolution equations~\cite{Balitsky:2015qba,Balitsky:2016dgz}.

A selection of processes that probe the WW and the dipole (denoted with DP) TMDs is given in Table~\ref{f1table}.~\cite{Boer:2016bfj}

\begin{table}[htb]
\centering
\caption{Processes that probe the WW and dipole gluon TMDs at small $x$. Table from Ref.~\protect\cite{Boer:2016bfj}.}
\label{f1table}       % Give a unique label
% For LaTeX tables you can use
\resizebox{12.7cm}{!}{
\begin{tabular}{|l|c|c|c|c|c|c|c|}
\hline
\hline
{}& {} & {} & {} & {} & {} &{} & {}\\[-2 mm]
{}& DIS & DY & SIDIS & $pA\to \gamma\, {\rm jet}\, X$ & $e \, p \to e'\, Q \, \overline{Q} \, X$ & $p p \to \eta_{c,b} \, X$ & $pp \to J/\psi\, \gamma\, X$ \\[0.4 mm]
{}& {} & {} & {} & {} & $e \, p \to e'\, j_1 \, j_2 \, X$ & $p p \to H \, X \hspace{2mm} $ & $pp \to \Upsilon\, \gamma\, X\hspace{3mm} $ \\[0.4mm]\hline
WW & $\times$ &$\times$ & $\times$ & $\times$ & $\surd$ & $\surd$ & $\surd$ \\[0.6 mm]\hline
DP & $\surd$ &$\surd$ & $\surd$ & $\surd$ & $\times$ & $\times$ & $\times$ \\[0.4 mm]\hline\hline
%| & first & second & third  \\\hline
%number & number & number \\
%number & number & number \\\hline
\end{tabular}
}
% Or use
%\vspace*{5cm}  % with the correct table height
\end{table}

\section{Results from the Connection Between the TMD Formalism and the CGC}
\label{sec:results}

In this section we review some recent results based on the equivalence between the TMD formalism and the CGC theory. Using the definitions of the TMDs in terms of CGC correlators of Wilson lines from Eq.~\ref{eq:fgg6} one can draw conclusions about their behavior at small $x$. We will present results for the TMDs in the GBW and MV models and their JIMWLK evolution. Finally, we will present an improved TMD factorization formula for dijet production that can be used for studying the onset of saturation effects in pA collisions and in UPC.

\subsection{Gluon TMDs in the GBW model} 
\begin{figure}[t]
\begin{center}
\includegraphics[width=0.55\textwidth]{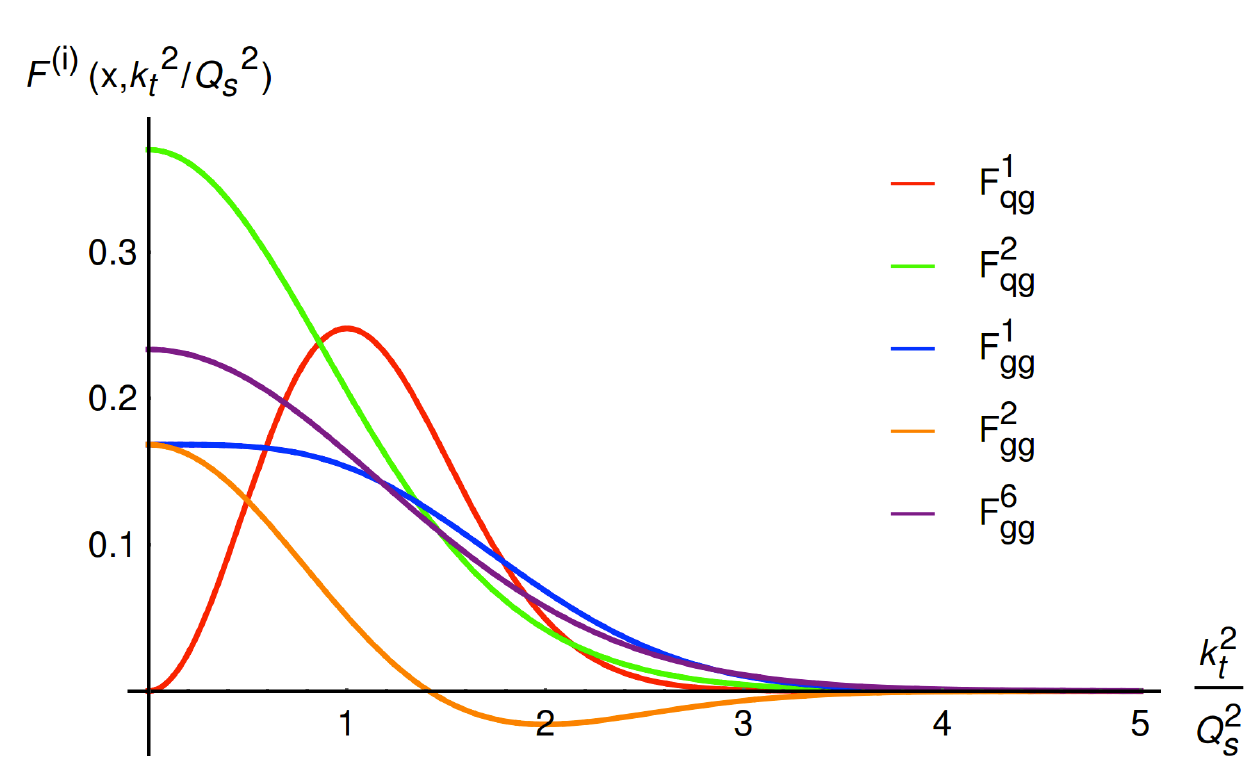}
\hfill
\includegraphics[width=0.38\textwidth]{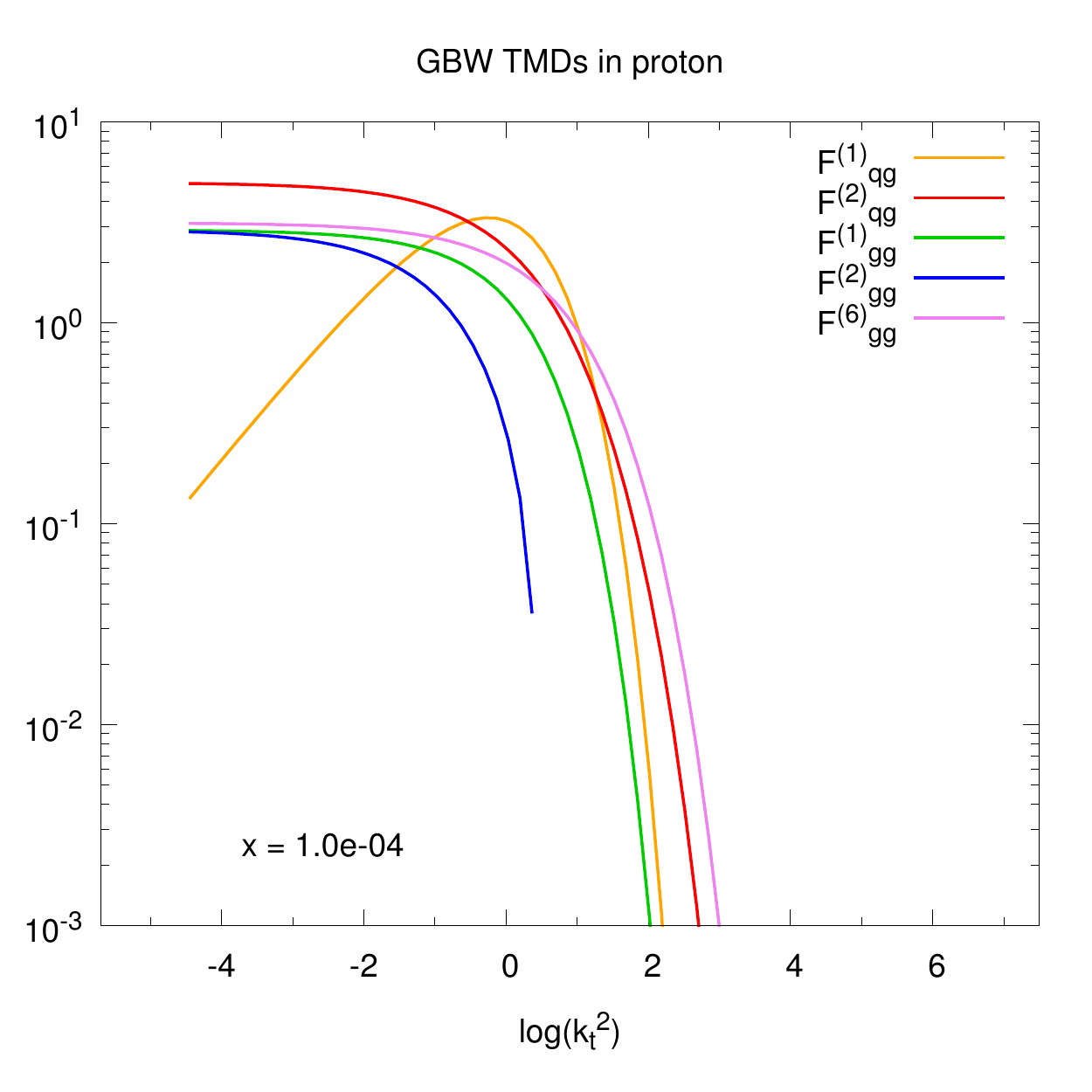}
 \end{center}
\caption{The gluon TMDs in the GBW model (up to a constant factor) as a function of $k_t^2/Q_s^2$ (left) and as a function of $\log (k^2_t/\mbox{GeV}^2)$ at $x=10^{-4}$ (right). Figure from Ref.~\protect\cite{vanHameren:2016ftb}.}
\label{fig:GBWgluons}
\end{figure}

The GBW model~\cite{GolecBiernat:1998js} is a phenomenological model for the dipole distribution that describes DIS data (in electron-proton collisions) at small $x$ and for moderate values of the photon virtuality. It gives an expression for $\mathcal{F}_{qg}^{(1)}$ at some fixed $x_0$, usually $x_0<0.01$:
\be 
\mathcal{F}_{qg}^{(1)}(x_2,k_t)\mid_{x_2=x_0} = 
\frac{N_c S_\perp}{2\pi^3 \alpha_s Q_s^2(x_2)} k_t^2 \exp\left[ -\frac{k_t^2}{Q_s^2(x_2)}\right]\ ,
\label{DipoleGBW}
\ee
where $S_\perp$ is the transverse area of the proton. The advantage of the GBW model is that one can derive analytical expressions for the distributions $\mathcal{F}_{qg}^{(2)}$, $\mathcal{F}_{gg}^{(1)}$, $\mathcal{F}_{gg}^{(2)}$, $\mathcal{F}_{gg}^{(3)}$ and $\mathcal{F}_{gg}^{(6)}$, based on their relation to the dipole distribution in the large $N_c$ limit~\cite{Dominguez:2010xd,Dominguez:2011wm,vanHameren:2016ftb}. The expressions are:~\cite{vanHameren:2016ftb}
\begin{eqnarray}
  {\cal F}_{qg}^{(2)}(x_2,k_t)\mid_{x_2=x_0}
  & = &
  \frac{N_c S_\perp}{4\pi^3 \alpha_s}
  \left[
    \Ei\left(-\frac{k_t^2}{Q_s^2(x_2)}\right) -
    \Ei\left(-\frac{k_t^2}{3Q_s^2(x_2)}\right)
  \right]\,, \nonumber
  \\
  {\cal F}_{gg}^{(1)}(x_2,k_t)\mid_{x_2=x_0}
  & = &
  \frac{N_c S_\perp}{16\pi^3 \alpha_s}
  \exp\left[-\frac{k_t^{2}}{2Q_s^2(x_2)}\right]\left(2 + \frac{k_t^{2}}{Q_s^2(x_2)} \right)\,, \nonumber
  \\
  {\cal F}_{gg}^{(2)}(x_2,k_t)\mid_{x_2=x_0}
  & = &
  \frac{N_c S_\perp}{16\pi^3 \alpha_s}
  \exp\left[-\frac{k_t^{2}}{2Q_s^2(x_2)}\right]\left(2 - \frac{k_t^{2}}{Q_s^2(x_2)} \right)\,, \nonumber
  \\
  {\cal F}_{gg}^{(3)}(x_2,k_t) \mid_{x_2=x_0}
  & = &
  \frac{N_c S_\perp}{4\pi^3 \alpha_s} \Ei\left(\frac{k_t^{2}}{2Q_s^2(x_2)}\right)\, , \nonumber 
  \\
  {\cal F}_{gg}^{(6)}(x_2,k_t) \mid_{x_2=x_0}
  & = &
  \frac{N_c S_\perp}{4\pi^3 \alpha_s}
  \left[
    \Ei\left(-\frac{k_t^2}{2Q_s^2(x_2)}\right) -
    \Ei\left(-\frac{k_t^2}{4Q_s^2(x_2)}\right)
  \right]\,. \label{FinalFgg6}
\end{eqnarray}
The exponential integral function is $\Ei(x) \equiv \int_x^\infty {dt}\, e^{-t}/t$, and ${\cal F}_{gg}^{(4)} = {\cal F}_{gg}^{(3)}$ in the GBW model. The behavior of the TMDs that follows from the GBW analytical expressions is plotted in Fig.~\ref{fig:GBWgluons} using $Q_s=0.88$ GeV at $x=10^{-4}$.~\cite{vanHameren:2016ftb}

The disadvantage of the GBW model is that it does not capture the right behavior of the TMDs at large $k_t$ for $k_t \gg Q_s$. 

\subsection{Gluon TMDs in the MV model} 

The right perturbative behavior of the TMDs at large $k_t$ can be obtained in the MV model. The MV model result for the dipole distribution is also independent of $x$, it is usually taken at $x_0 =0.01$ and serves as an initial condition for the small-$x$ evolution. In the Gaussian approximation, where the color field correlations are local, $\langle \rho_{\bf x} \rho_{\bf y} \rangle \sim \delta^{(2)}({\bf x} - {\bf y})$, the dipole distribution is:
\be 
\mathcal{F}_{qg}^{(1)}(x_2,k_t)\mid_{x_2=x_0} = 
\frac{N_c}{2\pi^2 \alpha_s} k_t^2 \int\frac{d^2{\bf x} d^2{\bf y}}{(2\pi)^2}\ e^{-ik_t\cdot({\bf x}-{\bf y})}
e^{-\frac{({\bf x} -{\bf y})^2 Q_s^2}{4}\ln \frac{1}{|{\bf x} -{\bf y}|\Lambda} },
\label{DipoleMV}
\ee
where $\Lambda$ is an infrared cutoff. The WW distribution is:
\be
\mathcal{F}_{gg}^{(3)}(x_2,k_t)= \frac{2C_F}{\pi^2 \alpha_s }\int\frac{d^2{\bf x} d^2{\bf y}}{(2\pi)^2}\
e^{-i{k_t}\cdot({\bf x}-{\bf y})}\ \frac{1}{({\bf x}-{\bf y})^2} \left[1 - e^{-\frac{C_A}{C_F} \frac{({\bf x} -{\bf y})^2 Q_s^2}{4}\ln \frac{1}{|{\bf x} -{\bf y}|\Lambda} }\right] \ ,
\label{eq:WW_AdjointDipole}
\ee
where $C_A$ is the Casimir of the adjoint representaion of $SU\left(N\right)$. The large $k_t$ behavior of the distributions can be obtained by expanding the exponential under the integral for small $|{\bf x} - {\bf y}|$. From this result one can derive the large $k_t$ behavior for some of the other TMDs in the large $N_c$ limit and obtain~\cite{vanHameren:2016ftb}:
\bea
\mathcal{F}_{qg}^{(1)}\ , \mathcal{F}_{qg}^{(2)}\ , \mathcal{F}_{gg}^{(1)}\ , \mathcal{F}_{gg}^{(6)} &=& 
\frac{N_c S_\perp Q_s^2}{4\pi^3 \alpha_s k_t^2} + \mathcal{O}\left(\frac{Q_s^4}{k_t^4}\log{\frac{k_t^2}{\Lambda^2}} \right) \ ,
\label{eq:highktbehavior}
\nonumber \\
\mathcal{F}_{gg}^{(2)} &=& 
\mathcal{O}\left(\frac{Q_s^4}{k_t^4}\log{\frac{k_t^2}{\Lambda^2}} \right) \ .
\eea

\begin{figure}
  \begin{center}
    \includegraphics[width=0.8\textwidth]{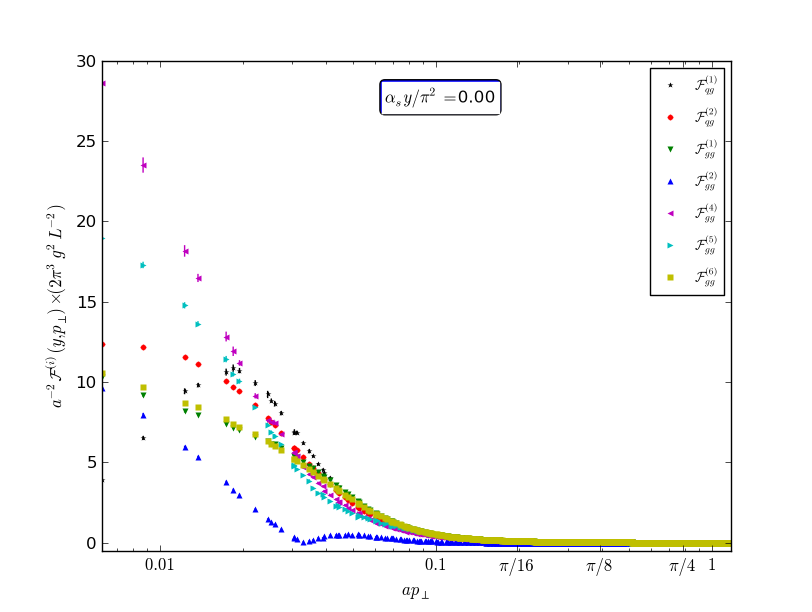}
  \end{center}
  \caption{TMDs in the MV model. Figure from Ref.~\protect\cite{Marquet:2016cgx}.}
  \label{fig:TMDs_0}
\end{figure}

In the MV model ${\cal F}_{gg}^{(4)} = {\cal F}_{gg}^{(3)}$. We see that all the gluon TMDs behave as $\sim Q_s^2/k_t^2$ for $k_t \gg Q_s$ except $\mathcal{F}_{gg}^{(2)}$ which vanishes to this order. The behavior of the TMDs in the MV model for any value of $k_t$ can be obtained numerically. Results from a lattice calculation of gluon TMDs in the MV model are given in Fig.~\ref{fig:TMDs_0} which confirms the asymptotic behavior derived analytically (the value of $y=0$ corresponds to the initial value $x_0$). In the limit of high-transverse momentum, the probe most probably resolves only one gluon in the hadronic or nuclear wave function with which it exchanges a large momentum. In this case, the gauge links in the TMD definitions drop out and all the TMDs fall on one universal curve.

\subsection{JIMWLK evolution of gluon TMDs}
\label{sub:JIMWLK}

\begin{figure}
  \begin{center}
    \includegraphics[width=0.8\textwidth]{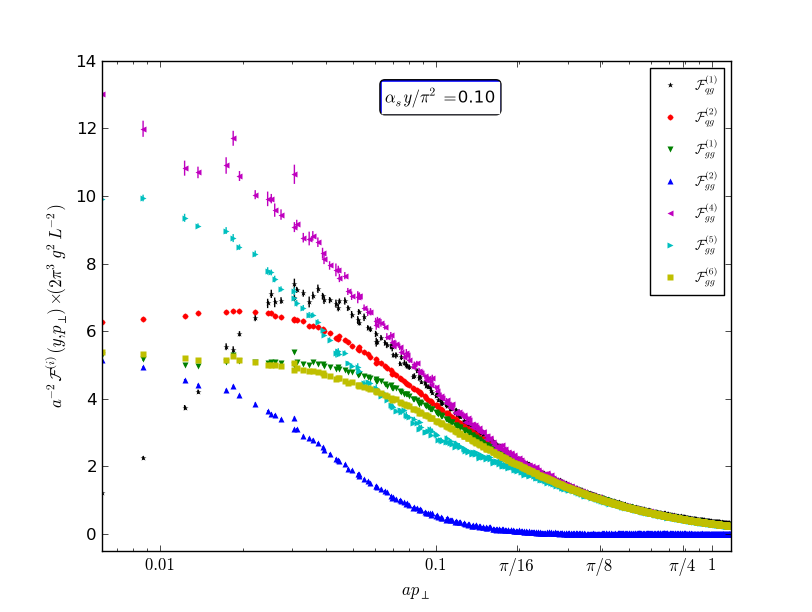}
    \includegraphics[width=0.8\textwidth]{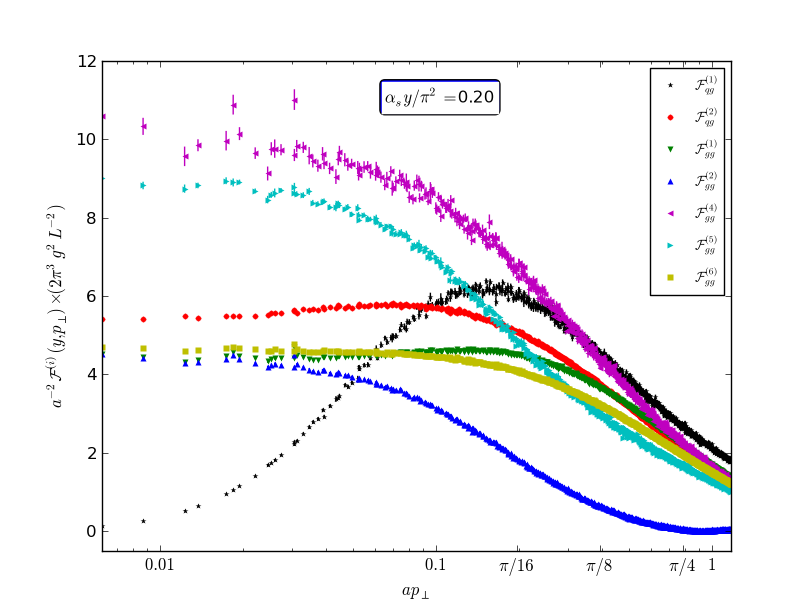}
  \end{center}
  \caption{Momentum dependence of gluon TMDs from JIMWLK evolution towards small $x$ from top to bottom plot. Figure from Ref.~\protect\cite{Marquet:2016cgx}.}
  \label{fig:TMDs_evolved}
\end{figure}

The JIMWLK evolution equation has been solved numerically~\cite{Blaizot:2002np,Rummukainen:2003ns,Dumitru:2011vk,Lappi:2012vw} and it has been used to study the behavior of the dipole and WW TMDs (see {\it e.g.}\ Ref.~\cite{Dumitru:2015gaa} for the WW TMD). Using the MV model as an initial condition (Fig.~\ref{fig:TMDs_0}), one can obtain the evolution of the TMDs for smaller values of $x$ by solving the JIMWLK equation on the lattice. Here, we present results from Ref.~\cite{Marquet:2016cgx} in order to compare the behavior of all the TMDs defined above. The results are given in Fig.~\ref{fig:TMDs_evolved} for two different steps of the small-$x$ evolution (the lower plot corresponds to smaller $x$). The small-$x$ evolution of ${\cal F}_{gg}^{(4)}$ with the MV initial condition is identical to the small-$x$ evolution of ${\cal F}_{gg}^{(3)}$. From the top plot in Fig.~\ref{fig:TMDs_evolved} we see that the universal behavior of the TMDs at high $k_t$, that is observed in the MV model, is preserved by the small-$x$ evolution: all the TMDs fall on the same curve again, with the difference that the fall-off changes from $1/k_t^2$ in the MV model to a less steep power $1/k_t^{2\gamma}$ with $\gamma < 1$, and it happens for larger values of $k_t$.

The behavior at small $k_t$ (in the saturation regime) is best seen in the lower plot of Fig.~\ref{fig:TMDs_evolved} for even smaller values of $x$. The TMDs at low $k_t$ behave differently, but the differences are under control in the CGC theory. These results for the evolution of the TMDs can be used in cross sections for other processes where they appear. 

Finally, from the small-$x$ evolution one can observe geometric scaling~\cite{Stasto:2000er} for all TMDs, {\it i.e.} after some evolution the TMDs depend on $k_t/Q_s(x_2)$ only, as opposed to $k_t$ and $x_2$ separately.~\cite{Marquet:2016cgx}. The geometric scaling of the TMDs is to be expected from the arguments presented in Ref.~\cite{Munier:2003vc}, and it was conjectured to hold for the WW distribution in Ref.~\cite{Dominguez:2011gc}. The geometric scaling from JIMWLK evolution was studied in Ref.~\cite{Lappi:2011ju} for the dipole distribution, and in Ref.~\cite{Dumitru:2015gaa} for the WW distribution, while Ref.~\cite{Marquet:2016cgx} observed geometric scaling for all the TMDs.

\subsection{Saturation effects from an improved TMD factorization}
\label{sub:ITMD}

In this last subsection we will discuss some applications of the TMD factorization framework to studying saturation effects in high-energy collisions. Measurements of forward di-hadron production in deuteron-gold (d+Au) collisions at the Relativistic Heavy Ion Collider (RHIC)\cite{Adare:2011sc,Braidot:2010zh} can be successfully explained by calculations in the CGC theory~\cite{Marquet:2007vb,Albacete:2010pg,Stasto:2011ru,Lappi:2012nh}, which can be considered as a form of evidence of the saturation of gluon densities.\footnote{For other explanations, not based on the saturation formalism, see {\it e.g.}\ Refs.~\cite{Qiu:2004da,Kang:2011bp}.} At the Large Hadron Collider (LHC) one can study the same process while taking into account the large transverse momenta of the final state partons, which provide a hard scale in the problem, much larger than the saturation scale. When the jets propagate almost back-to-back in the transverse plane, their momentum imbalance is sensitive to saturation effects in the nucleus. Instead of the full CGC cross section one can therefore use its correlation limit, {\it i.e.}\ the TMD factorized form which is valid when the momenta are ordered, $Q_s \sim k_t \ll P_t$. Ref.~\cite{Stasto:2011ru} studied di-hadron correlations in d+Au collisions using the effective TMD factorization formula derived in Ref.~\cite{Dominguez:2011wm} in the large $N_c$ limit, while Ref.~\cite{Zheng:2014vka} studied di-hadron correlations in DIS with a nucleus in the correlation limit using Eq.~\ref{eq:dijet_DIS}. These formulas, however, only capture correctly the small-$k_t$ region, for $k_t \ll P_t$.

The limit of large $k_t$ of the order of the hard scale $P_t$, $k_t \sim P_t$, is the region where the HEF framework is applicable.~\cite{Catani:1990eg,Deak:2009xt} The HEF formula for forward dijet production in pA collisions involves collinear PDFs describing the proton, off-shell matrix elements describing the hard part and one unintegated gluon distribution describing the nucleus in the small $x$ regime. Unlike in the TMD formalism, in the HEF cross section the $k_t$ dependence survives in the matrix elements and they are calculated with an off-shell gluon from the nucleus. The HEF formula has been used to study saturation effects in the nucleus in Refs.~\cite{vanHameren:2014ala,Kutak:2012rf,vanHameren:2014lna,vanHameren:2013fla}. Similarly to TMD factorization, HEF is also contained in the CGC cross section, now as its dilute limit. Even though the CGC theory is constructed for the saturation region, it captures the correct perturbative limit outside of the saturation region and it can be applied for dilute systems, provided that the small-$x$ limit is considered.

The dilute limit of the CGC corresponds to the large $k_t$ limit, $k_t \gg Q_s$. Physically, the rescatterings of the incoming partons off the gluons in the hadronic or nuclear wave function, in this limit, are reduced to one hard scattering off one gluon, as discussed above for the large $k_t$ behavior of the TMDs in the MV model. In this single scattering approximation, the Wilson lines in the correlators in the CGC amplitude squared can be expanded to second order in the background field (two-gluon exchange). The cross section in this limit involves only the dipole distribution.~\cite{Iancu:2013dta,Kotko:2015ura}. It can be shown that the dilute limit of the CGC cross section for forward dijet production is equivalent to the HEF formula and a relation between the dipole distribution and the unintegrated distribution in HEF can be established.~\cite{Kotko:2015ura} The dipole distribution in the dilute approximation is not sensitive to non-linear effects and evolves according to the BFKL equation. The dilute limit has been considered for other processes in pA collisions as well, {\it e.g.}\ for forward jet production~\cite{Dumitru:2002qt}, Higgs production~\cite{Schafer:2012yx}, inclusive photon production at next-to-leading order~\cite{Benic:2016uku} and dijet plus photon production~\cite{Altinoluk:2018uax}. 

The full CGC cross section captures correctly both limits of $k_t\sim Q_s$ and $k_t\sim P_t$, and the region in between, but a factorized formula is more amenable for phenomenological applications. It is therefore preferable to have a factorization that can describe the whole region of $k_t$ values such that one can observe the onset of saturation, {\it i.e.}\ the transition from large $k_t$, where saturation effects are negligible, to the region of $k_t\sim Q_s$, where saturation effects are most pronounced. This can be achieved with a model, called ITMD, which is a generalization of the effective TMD factorization in section~\ref{sec:TMD}, now with off-shell matrix elements, and which interpolates between the TMD cross section at small $k_t$ and the HEF cross section at large $k_t$.~\cite{Kotko:2015ura} In the appropriate limits the ITMD formula reduces to the TMD or HEF factorizations. 

\begin{figure}
\centering
\begin{minipage}{.5\textwidth}
  \centering
  \includegraphics[width=.85\linewidth]{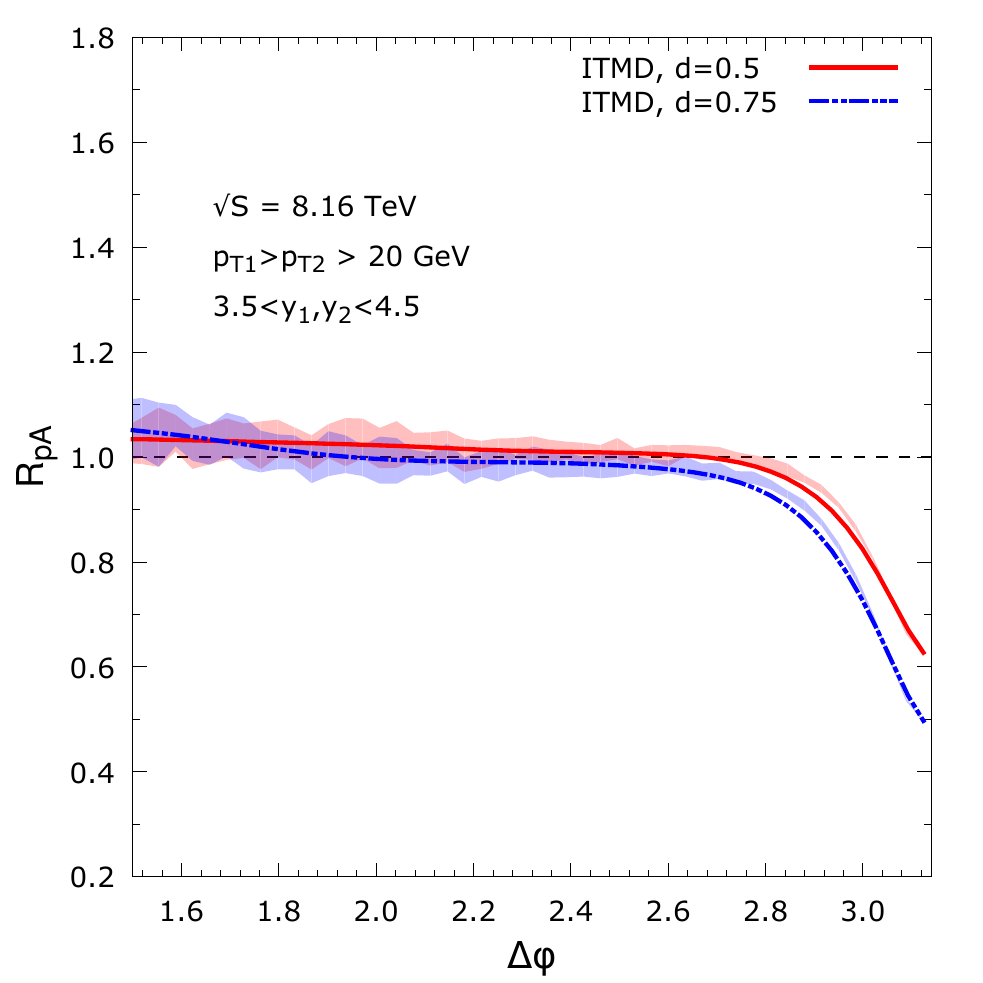}
  %\captionof{figure}{Nuclear modification factor from ITMD factorization for forward dijet production in pA {\it vs.}\ pp collisions. Figure from Ref.~\protect\cite{vanHameren:2016ftb}.}
  %\label{fig:pA}
\end{minipage}%
\begin{minipage}{.5\textwidth}
  \centering
  %\vspace{-0.cm}
  \includegraphics[width=.85\linewidth]{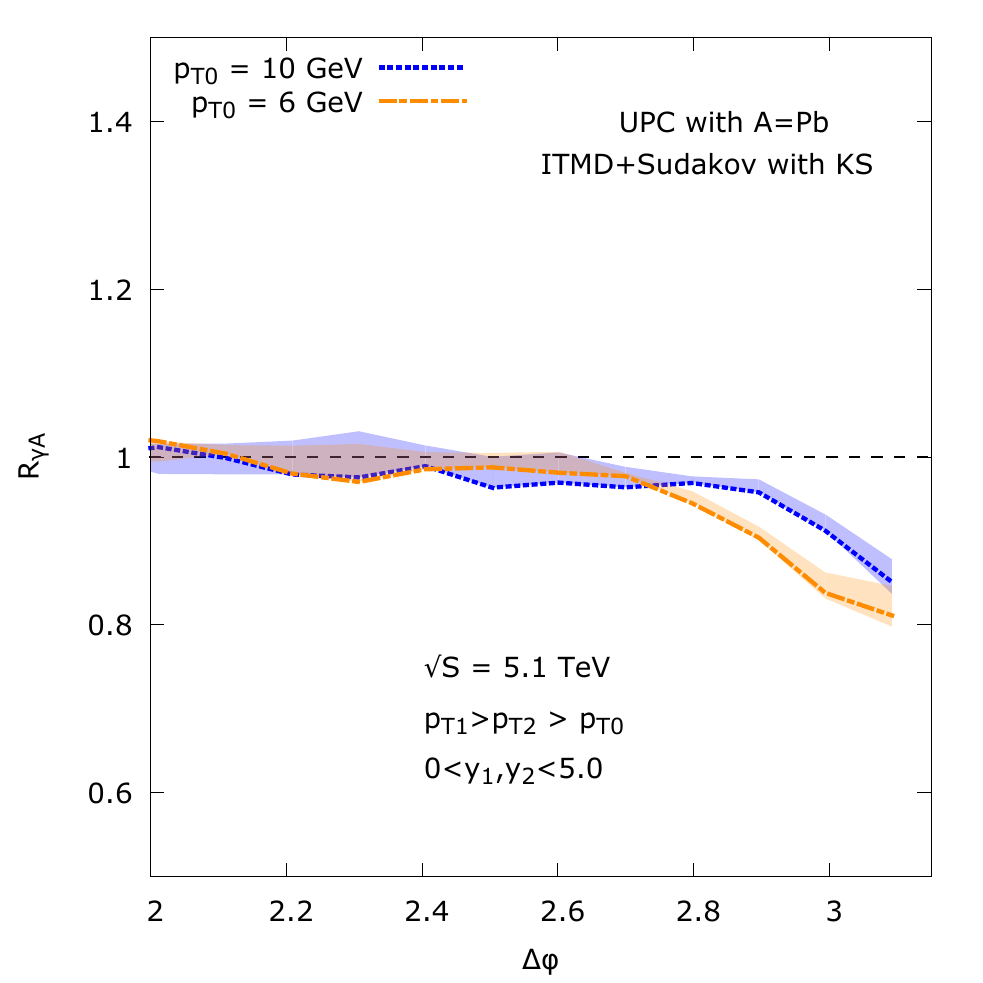}
  %\captionof{figure}{Nuclear modification factor from ITMD factorization for forward dijet production in UPC {\it vs.}\ Ap collisions. Figure from Ref.~\protect\cite{Kotko:2017oxg}.}
 % \label{fig:upc} 
\end{minipage}
\caption{Left plot: Nuclear modification factor from the ITMD framework for forward dijet production in pA {\it vs.}\ pp collisions. Figure from Ref.~\protect\cite{vanHameren:2016ftb}. Right plot: Nuclear modification factor from the ITMD framework for forward dijet production in UPC {\it vs.}\ Ap collisions. Figure from Ref.~\protect\cite{Kotko:2017oxg}.}
\label{fig:pA_upc}
\end{figure}

Ref.~\cite{vanHameren:2016ftb} used an ITMD factorization with five gluon TMDs (in the large $N_c$ limit) describing the nucleus to study saturation effects in forward dijet production in pA collisions. Ref.~\cite{Kotko:2017oxg} used an ITMD factorization which involved only the WW gluon TMD to study saturation effects in forward dijet production in ultra-peripheral heavy ion collisions (UPC) from a photon initiated process, $\gamma + A \to 2 {\text jet} + X$, and used a model to estimate the effects of the Sudakov logarithms (see the next section). We show the results for the nuclear modification factors from these works in Fig.~\ref{fig:pA_upc}. 

The nuclear modification factor, defined as:
\begin{equation}
R_{\rm pPb} = \frac{\displaystyle \frac{d\sigma^{p+Pb}}{d{\cal O}}}
                 {\displaystyle A\ \frac{d\sigma^{p+p}}{d{\cal O}}}\,,
\label{eq:RpA}
\end{equation}
with $A=208$ for the lead nucleus, as a function of the angle $\Delta \phi$ between the final jets, from Ref.~\cite{vanHameren:2016ftb}, is shown in the left plot in Fig.~\ref{fig:pA_upc}. The nuclear modification factor, defined as:
\begin{equation}
R_{\rm \gamma A} = \frac{\displaystyle {d\sigma^{UPC}_{AA}}}
                 {\displaystyle A\ {d\sigma^{UPC}_{Ap}}}\,,
\label{eq:upc}
\end{equation}
with the same $A=208$,  as a function of the angle $\Delta \phi$ between the final jets, from Ref.~\cite{Kotko:2017oxg}, is shown in the right plot in Fig.~\ref{fig:pA_upc}.{\footnote{In both results the TMDs are taken from the Kutak-Sapeta (KS) solution~\cite{Kutak:2012rf} of the BK equation, which accounts for higher-order corrections. The same analysis with the full JIMWLK evolution of the TMDs from subsection~\ref{sub:JIMWLK} can be performed in the future.} }If saturation effects are not present the nuclear modification factor should be equal to one. From the plots in Fig.~\ref{fig:pA_upc} one can observe an onset of gluon saturation around $\Delta \phi = \pi$, {\it i.e.}\ in the correlation limit when the jets are almost back-to-back. Away from this region, when the momentum imbalance is large, the process is not sensitive to the non-linear effects in the nucleus, and the value of the nuclear modification factor is close to one.

The phenomenological applications of the ITMD approach discussed here do not incorporate contributions from Sudakov double logarithms or use a model for the Sudakov resummation. The Sudakov logarithms are important for reliable phenomenological predictions and we discuss their implementation in the next section.

\section{Unifying the TMD evolution and the non-linear small-$x$ evolution}
\label{sec:evolution}

In this section we review some recent results on unifying the TMD evolution, obtained by resumming large Sudakov logarithms, with the non-linear small-$x$ evolution of gluon TMDs.{\footnote{For the combined evolution of the quark TMD see Ref.~\cite{Kovchegov:2015zha}.}} TMD factorization formulas are derived for processes where there exists ordering of momentum scales. In the previous sections we discussed the correlation limit for different cases, for example, for the limit of small momentum imbalance of two jets produced in pA collisions compared to their individual transverse momenta, $k_t \ll P_t$. The cross sections for these kind of processes receive large contributions from Sudakov double logarithms, $ \ln^2 Q^2/k_t^2$, and thus terms of the type $\alpha_s\ln^2 Q^2/k_t^2$ need to be resummed to all orders. This is achieved by the Collins-Soper-Sterman resummation~\cite{Collins:1985}. In order to have a reliable theory at small $x$ and in the correlation limit one needs to consistently resum small-$x$ logarithms, $\alpha_s \ln\, 1/x$, and Sudakov double logarithms, $\alpha_s\ln^2 Q^2/k_t^2$, in parallel. 

The Sudakov double logarithms, at the level of the cross section, were resummed in the saturation framework in Refs.~\cite{Mueller:2012uf,Mueller:2013wwa} for various processes from one loop calculations.\footnote{For an earlier work see Ref.~\cite{Levin:2010zs}.} The results can be summarized schematically as:
\be 
\frac{d\sigma}{dy_1dy_2d^2P_td^2k_t} \propto
H\left( P_t^2 \right)\, \int d^2 {\bf x} d^2 {\bf y} 
\ e^{i k_t \cdot \left({\bf x} -{\bf y}\right)}
\, e^{-\mathcal{S}_{sud}\left(P_t, {\bf R}\right)}
\mathcal{W}_{x_2}\left({\bf x},{\bf y}\right) \, ,
\ee
where $P_t$ is the hard and $k_t$ the soft scale, and ${\bf R} \equiv {\bf x} - {\bf y}$. The hard perturbative part is denoted with $H$. The correlators of Wilson lines that define the unintegrated gluon distributions for the nucleus are denoted with $\mathcal{W}_{x_2}$. The small-$x$ logarithms are resummed through JIMWLK or BK evolution equations of $\mathcal{W}_{x_2}$. The Sudakov double logarithms are resummed in the exponential of the Sudakov factor:
\be 
{S}_{sud} = \frac{\alpha_s}{\pi} \mathcal {C} 
\int_{c_0^2/{\bf R}^2}^{P_t^2} \frac{d \mu^2}{\mu^2} \ln \frac{P_t^2}{\mu^2} \, ,
\ee 
where $c_0$ is a parameter of order one, and $\mathcal{C}$ is a coefficient that needs to be calculated for each process individually and is determined by the color structure of the particular process. The Sudakov factors for Higgs boson production in pA collisions, heavy quark pair production and dijet production in DIS, jet-photon production and dijet production in pA collisions were calculated in Ref.~\cite{Mueller:2013wwa}. The Sudakov resummation and the small-$x$ evolution were combined, for instance, in Ref.~\cite{Watanabe:2015yca} to interpret the data for forward $J/\Psi$ and $\Upsilon$ productions at the LHC. The results from Ref.~\cite{Mueller:2013wwa} were used in Ref.~\cite{Altinoluk:2018uax} to estimate the effects of the Sudakov logarithms in the process of dijet plus soft photon production in pA collisions and it was found that they may wash away the effects of saturation, but not completely.

The TMD evolution and the small-$x$ evolution of the TMDs with future pointing gauge links, $\mathcal{F}_{gg}^{(3)}$, and past pointing gauge links, $\mathcal{F}_{gg}^{(4)}$, were discussed in Ref.~\cite{Balitsky:2015qba} and Ref.~\cite{Balitsky:2016dgz}, respectively, where the authors derived evolution equations valid for any $x$ and any $k_t$. We refer the reader to those references for the full expressions of the evolution equations. We point out that the full evolution equations for $\mathcal{F}_{gg}^{(3)}$ and $\mathcal{F}_{gg}^{(4)}$ are different. In the appropriate limits the equations reduce to known results. In the small-$x$ limit both of the evolution equations are the same and coincide with the small-$x$ nonlinear evolution equation for the WW distribution~\cite{Dominguez:2011gc}. For moderate $x$ values, but small $k^2_t \ll s$, the evolution equations reduce to a linear evolution equation of the Sudakov type with resummed Sudakov double logarithms.

In Ref.~\cite{Xiao:2017yya} the authors resummed the small-$x$ and Sudakov logarithms simultaneously for the WW ($\mathcal{F}_{gg}^{(3)}$) and dipole ($\mathcal{F}_{qg}^{(1)}$) distributions, in the small-$x$ region. Their procedure is similar to the resummation performed at the level of the cross section in Refs.~\cite{Mueller:2012uf,Mueller:2013wwa}, but with the difference that now both resummations are done at the level of the TMDs, without considering a particular scattering process. The results can be applied to any process where the dipole and WW TMDs appear. We refer the reader to Ref.~\cite{Xiao:2017yya} for the explicit expressions. 

\section{Conclusions}
\label{sec:conclusions}

In this contribution we reviewed some aspects of two independent frameworks to describe high-energy scatterings, namely the TMD factorization approach and the CGC effective field theory, focusing the discussion on their common region of applicability at small $x$ and on some recent results on gluon saturation based on their connection. In particular, we recalled the origin of non-universality of gluon TMDs in the two frameworks and we discussed how some universal properties can be recovered in the CGC theory. The non-universality of gluon TMDs is the result of their process dependence; different types of color interactions (initial, final or both) present in different processes determine the gauge link structure of the TMDs, making them gauge invariant. Understanding the process dependence of TMDs is important for determining which distribution is being extracted from a particular process and for making an exact connection between experiments which probe different TMDs. 

Measurements of TMDs provide information on the three-dimensional momentum structure of the proton or nucleus, and at small $x$ also on the phenomenon of gluon saturation. Extractions of gluon TMDs from a global data analysis is desirable in the future. The WW gluon TMD is the proper number density of gluons in the hadronic or nuclear wave function and it was proposed recently that it can be measured in dijet production in DIS. It was later identified in other processes as well (see Table~\ref{f1table}). Such measurements can be performed at the planned EIC and direct information on gluon saturation can be obtained. In the saturation framework, the dipole gluon distribution of the proton is the one that is the most constrained by DIS data from the Hadron-Electron Ring Accelerator (HERA), and has been used, for instance, in both single inclusive hadron production and photon-jet correlations in pp and pA collisions. The future EIC is needed for a measurement of the nuclear dipole TMD and for an extraction of the nuclear saturation scale. More complicated processes involve both fundamental distributions along with other TMDs. We reviewed the process of dijet production in pA collisions which involves eight different gauge link structures and which can be used to study the onset of gluon saturation at the LHC.

The process dependence of TMDs is a consequence of the gauge invariance of QCD that can be traced explicitly, and in that way tested in the connection between experiments. In the opposite direction, the process dependence of TMDs can be used as a test of the computational frameworks applied to a particular case. Processes with the same color structure of the hard part should involve the same gluon TMDs. For example, the full CGC cross section for dijet plus photon production in pA collisions involves the same gluon TMDs in the correlation limit as the CGC cross section for dijet production in pA collisions, confirming the accuracy of the former one.

We also reviewed the connection between the CGC theory and the TMD and HEF frameworks. The main conclusion is that, at small $x$, the CGC theory captures both of these factorization formalisms. When an ordering of momentum scales is imposed, the full CGC cross section reduces to a TMD factorized form, while when the dilute limit is considered, the full CGC cross section reduces to the HEF form. Based on this connection one can study the properties of gluon TMDs at small $x$ using saturation models and their JIMWLK evolution. In addition, one can study saturation effects in high-energy collisions with a model formula (ITMD) that interpolates between the TMD and HEF limits, instead of the more complicated CGC cross section. We discussed results based on the ITMD framework for dijet production in pA collisions and in UPC; similar investigations can be performed for other processes in the future.

One of the main lines of research at the moment is unifying the TMD evolution with the non-linear small-$x$ evolution. We reviewed some recent progress along these lines, which is mainly focused on the WW and dipole gluon distribution. For improved phenomenological studies of saturation effects the implementation of Sudakov resummation needs to be extended to the other TMDs as well.

In summary, a combined framework of CGC theory and TMD factorization can provide us with a better understanding of the three-dimensional momentum structure of protons and nuclei, of the phenomenon of gluon saturation at small $x$ and of QCD properties of high-energy collisions in general.

\section*{Acknowledgements}

I would like to thank all of my collaborators from the projects covered in this review: Tolga Altinoluk, N\'{e}stor Armesto, Andreas van Hameren, Piotr Kotko, Alex Kovner, Krzysztof Kutak, Michael Lublinsky, Cyrille Marquet, Claude Roiesnel and Sebastian Sapeta. I thank Piet Mulders and Dani\"el Boer for many clarifying discussions on the subjects presented in this review. I thank Tom van Daal for his detailed reading of the draft and for his very useful feedback and discussions. I would like to thank the Nuclear Theory Group at the Brookhaven National Lab, where part of this review was presented, for their invitation and hospitality. I acknowledge support by the European Community under the ``Ideas'' programme QWORK (contract no. 320389).


\begin{thebibliography}{0}
\bibitem{1} G. Altarelli and G. Parisi, {\it Nucl.\ Phys.\ B} {\bf 126} 298 (1977).

\bibitem{2} Y. L. Dokshitzer, {\it Sov.\ Phys.\ JETP} {\bf 46}, 641 (1977) [{\it {Zh.\ Eksp.\ Teor.\ Fiz.\ }}  {\bf 73}, 1216 (1977)].

\bibitem{3} V. N. Gribov and L. N. Lipatov, {\it Sov.\ J.\ Nucl.\ Phys.\ } {\bf 15}, 438 (1972) [{\it {Yad.\ Fiz.\ }}  {\bf 15}, 781 (1972)].

\bibitem{4} L. N. Lipatov, {\it Sov.\ J.\ Nucl.\ Phys.\ } {\bf 20}, 94 (1975) [{\it {Yad.\ Fiz.\ }}  {\bf 20}, 181 (1974)].

\bibitem{Ball:2017nwa} 
  R. D. Ball {\it et al.} [NNPDF Collaboration],
  %``Parton distributions from high-precision collider data,''
 {\it Eur.\ Phys.\ J.\ C }{\bf 77}, no. 10, 663 (2017).
 
\bibitem{Dulat:2015mca} 
  S. Dulat {\it et al.},
  %``New parton distribution functions from a global analysis of quantum chromodynamics,''
  {\it Phys.\ Rev.\ D }{\bf 93}, no. 3, 033006 (2016).
  
\bibitem{Harland-Lang:2014zoa} 
  L. A. Harland-Lang, A. D. Martin, P. Motylinski and R. S. Thorne,
  %``Parton distributions in the LHC era: MMHT 2014 PDFs,''
  {\it Eur.\ Phys.\ J.\ C }{\bf 75}, no. 5, 204 (2015).
  
\bibitem{Butterworth:2015oua} 
  J. Butterworth {\it et al.},
  %``PDF4LHC recommendations for LHC Run II,''
 {\it J.\ Phys.\ G }{\bf 43}, 023001 (2016).    

 
\bibitem{Collins:CUP2011}
J. Collins, {\it Foundations of perturbative QCD }(Cambridge University Press, 2013). 

\bibitem{Collins:1981uw} 
  J. C. Collins and D. E. Soper,
  %``Parton Distribution and Decay Functions,''
  {\it Nucl.\ Phys.\ B }{\bf 194}, 445 (1982).
  
\bibitem{Collins:1981va} 
  J. C. Collins and D. E. Soper,
  %``Back-To-Back Jets: Fourier Transform from B to K-Transverse,''
  {\it Nucl.\ Phys.\ B }{\bf 197}, 446 (1982).  
  
\bibitem{Collins:1982wa} 
  J. C. Collins, D. E. Soper and G. F. Sterman,
  %``Factorization for One Loop Corrections in the {Drell-Yan} Process,''
 {\it Nucl.\ Phys.\ B }{\bf 223}, 381 (1983).    
  
\bibitem{Collins:1985}
  J. C. Collins, D. E. Soper and G. F. Sterman,
  %``Transverse Momentum Distribution in Drell-Yan Pair and W and Z Boson Production,''
 {\it Nucl.\ Phys.\ B} {\bf 250} 199 (1985). 
%  doi:10.1016/0550-3213(85)90479-1
  %%CITATION = doi:10.1016/0550-3213(85)90479-1;%%
  
  
  
%\bibitem{Collins:1981uk} 
%  J. C. Collins and D. E. Soper,
%  %``Back-To-Back Jets in QCD,''
% {\it Nucl.\ Phys.\ B }{\bf 193}, 381 (1981).

%\bibitem{6}
%  J. w. Qiu and X. f. Zhang,
%  %``QCD prediction for heavy boson transverse momentum distributions,''
%  {\it Phys.\ Rev.\ Lett.\ } {\bf 86} 2724 (2001).
%%  doi:10.1103/PhysRevLett.86.2724
% % [hep-ph/0012058].
%  %%CITATION = doi:10.1103/PhysRevLett.86.2724;%%
%
%\bibitem{7}
%  J. P. Ralston and D. E. Soper,
%  %``Production of Dimuons from High-Energy Polarized Proton Proton Collisions,''
%  {\it Nucl.\ Phys.\ B } {\bf 152} 109 (1979).
%%  doi:10.1016/0550-3213(79)90082-8
%  %%CITATION = doi:10.1016/0550-3213(79)90082-8;%%
%

%
%\bibitem{9}
%  L. N. Lipatov,
%  %``The Bare Pomeron in Quantum Chromodynamics,''
%  {\it Sov.\ Phys.\ JETP } {\bf 63} 904 (1986)
%   [{\it Zh.\ Eksp.\ Teor.\ Fiz.\  } {\bf 90} 1536 (1986)].
%  %%CITATION = SPHJA,63,904;%%
%
%\bibitem{10}
%  S. Catani, M. Ciafaloni and F. Hautmann,
%  %``High-energy factorization and small x heavy flavor production,''
%  {\it Nucl.\ Phys.\ B {\bf 366} } 135 (1991).
%  %%CITATION = NUPHA,B366,135;%%

\bibitem{Ji:2004wu} 
  X. d. Ji, J. p. Ma and F. Yuan,
  %``QCD factorization for semi-inclusive deep-inelastic scattering at low transverse momentum,''
  Phys.\ Rev.\ D {\bf 71}, 034005 (2005)

\bibitem{Bacchetta:2017gcc} 
  A. Bacchetta, F. Delcarro, C. Pisano, M. Radici and A. Signori,
  %``Extraction of partonic transverse momentum distributions from semi-inclusive deep-inelastic scattering, Drell-Yan and Z-boson production,''
  {\it JHEP }{\bf 1706}, 081 (2017).

%\bibitem{11}
%  P. J. Mulders and J. Rodrigues,
%  %``Transverse momentum dependence in gluon distribution and fragmentation functions,''
%  {\it Phys.\ Rev.\ D {\bf 63} } 094021 (2001).
%  doi:10.1103/PhysRevD.63.094021
%  [hep-ph/0009343].
  %%CITATION = doi:10.1103/PhysRevD.63.094021;%%
  
\bibitem{Collins:1983pk} 
  J. C. Collins, D. E. Soper and G. F. Sterman,
  %``Relation of Parton Distribution Functions in {Drell-Yan} Process to Deeply Inelastic Scattering,''
 {\it Phys.\ Lett.\ } {\bf 126B}, 275 (1983).  
 
 
\bibitem{Brodsky:2002cx} 
  S. J. Brodsky, D. S. Hwang and I. Schmidt,
  %``Final state interactions and single spin asymmetries in semiinclusive deep inelastic scattering,''
  {\it Phys.\ Lett.\ B }{\bf 530}, 99 (2002).  
  
\bibitem{Collins:2002kn} 
  J. C. Collins,
  %``Leading twist single transverse-spin asymmetries: Drell-Yan and deep inelastic scattering,''
  {\it Phys.\ Lett.\ B }{\bf 536}, 43 (2002).

\bibitem{12}
  A. V. Belitsky, X. Ji and F. Yuan,
  %``Final state interactions and gauge invariant parton distributions,''
  {\it Nucl.\ Phys.\ B } {\bf 656} 165 (2003).
%  [hep-ph/0208038].
  %%CITATION = HEP-PH/0208038;%%
  
\bibitem{Bomhof:2004aw} 
  C. J. Bomhof, P. J. Mulders and F. Pijlman,
  %``Gauge link structure in quark-quark correlators in hard processes,''
  {\it Phys.\ Lett.\ B }{\bf 596}, 277 (2004).  
  
\bibitem{Bomhof:2006dp} 
  C. J. Bomhof, P. J. Mulders and F. Pijlman,
  %``The Construction of gauge-links in arbitrary hard processes,''
 {\it Eur.\ Phys.\ J.\ C }{\bf 47}, 147 (2006).    
 
\bibitem{Boer:1999si} 
  D. Boer and P. J. Mulders,
  %``Color gauge invariance in the Drell-Yan process,''
 {\it Nucl.\ Phys.\ B }{\bf 569}, 505 (2000).  
  
\bibitem{13} 
  E. A. Kuraev, L. N. Lipatov and V. S. Fadin,
  %``The Pomeranchuk Singularity in Nonabelian Gauge Theories,''
  {\it Sov.\ Phys.\ JETP } {\bf 45}, 199 (1977)
  [{\it Zh.\ Eksp.\ Teor.\ Fiz.\  } {\bf 72}, 377 (1977)].  
  
\bibitem{14} 
  I. I. Balitsky and L. N. Lipatov,
  %``The Pomeranchuk Singularity in Quantum Chromodynamics,''
  {\it Sov.\ J.\ Nucl.\ Phys.\  } {\bf 28}, 822 (1978)
  [{\it Yad.\ Fiz.\  }{\bf 28}, 1597 (1978)].  
  
\bibitem{15} 
  L. V. Gribov, E. M. Levin and M. G. Ryskin,
  %``Semihard Processes in QCD,''
  {\it Phys.\ Rept.\  } {\bf 100}, 1 (1983).  
  
\bibitem{16} 
  A. H. Mueller and J. w. Qiu,
  %``Gluon Recombination and Shadowing at Small Values of x,''
  {\it Nucl.\ Phys.\ B } {\bf 268}, 427 (1986).  
  
\bibitem{17} 
  L. D. McLerran and R. Venugopalan,
  %``Computing quark and gluon distribution functions for very large nuclei,''
  {\it Phys.\ Rev.\ D } {\bf 49}, 2233 (1994).
  
\bibitem{18} 
  L. D. McLerran and R. Venugopalan,
  %``Gluon distribution functions for very large nuclei at small transverse momentum,''
  {\it Phys.\ Rev.\ D } {\bf 49}, 3352 (1994).    
  
\bibitem{19} 
  E. Iancu, A. Leonidov and L. McLerran,
  %``The Color glass condensate: An Introduction,''
  hep-ph/0202270.  
  
\bibitem{20} 
  E. Iancu and R. Venugopalan,
  %``The Color glass condensate and high-energy scattering in QCD,''
  In *Hwa, R.C. (ed.) et al.: Quark gluon plasma* 249-3363.  
  
\bibitem{21} 
  J. Jalilian-Marian and Y. V. Kovchegov,
  %``Saturation physics and deuteron-Gold collisions at RHIC,''
  {\it Prog.\ Part.\ Nucl.\ Phys.\  } {\bf 56}, 104 (2006).  
  
\bibitem{22}
  F. Gelis, E. Iancu, J. Jalilian-Marian and R. Venugopalan,
  %``The Color Glass Condensate,''
  {\it Ann.\ Rev.\ Nucl.\ Part.\ Sci.\  } {\bf 60} 463 (2010).
%  [arXiv:1002.0333 [hep-ph]].
  %%CITATION = ARXIV:1002.0333;%%  
  
\bibitem{Kharzeev:2003wz} 
  D. Kharzeev, Y. V. Kovchegov and K. Tuchin,
  %``Cronin effect and high p(T) suppression in pA collisions,''
 {\it Phys.\ Rev.\ D }{\bf 68}, 094013 (2003).  
 
\bibitem{Dominguez:2010xd} 
  F. Dominguez, B. W. Xiao and F. Yuan,
  %``$k_t$-factorization for Hard Processes in Nuclei,''
  {\it Phys.\ Rev.\ Lett.\  }{\bf 106}, 022301 (2011).   
 
\bibitem{Dominguez:2011wm}
  F. Dominguez, C. Marquet, B. W. Xiao and F. Yuan,
  %``Universality of Unintegrated Gluon Distributions at small x,''
  {\it Phys.\ Rev.\ D } {\bf 83}, 105005 (2011).   
  
\bibitem{JIMWLK1} J. Jalilian Marian, A. Kovner, A.Leonidov and H.
Weigert,
  %``The BFKL equation from the Wilson renormalization group,''
{\it Nucl. Phys. B }  {\bf  504} 415 (1997).

\bibitem{JIMWLK2} J. Jalilian Marian, A. Kovner, A.Leonidov and H.
%``The Wilson renormalization group for low x physics: Gluon evolution at
  %finite parton density,''
{\it Phys. Rev. D}   {\bf 59} 014014 (1999).

\bibitem{JIMWLK3}
J. Jalilian Marian, A. Kovner and H. Weigert,
% ``The Wilson renormalization group for low x physics: Gluon evolution at
  %finite parton density,''
{\it Phys. Rev. D }  {\bf 59} 014015 (1999).

\bibitem{JIMWLK4}
 A. Kovner and J.G. Milhano,
%``Vector potential versus color charge density in low-x evolution,''
{\it Phys. Rev. D }  {\bf 61} 014012 (2000).

\bibitem{JIMWLK5}
A. Kovner, J.G. Milhano and H. Weigert,
 %``Relating different approaches to nonlinear QCD evolution at finite  gluon
  %density,''
{\it Phys. Rev. D}  {\bf 62} 114005 (2000).

\bibitem{JIMWLK6}
 H. Weigert,
 %``Unitarity at small Bjorken x,''
  {\it Nucl. Phys. A}  {\bf  703} 823 (2002).
  
\bibitem{balitsky1} I. Balitsky,
%``Operator expansion for high-energy scattering,''
{\it Nucl. Phys. B}   {\bf 463}, 99 (1996).
%
\bibitem{balitsky2} I. Balitsky,
%``Factorization for high-energy scattering,''
{\it Phys. Rev. Lett.} {\bf 81} 2024 (1998).

\bibitem{balitsky3} I. Balitsky,
%``Factorization and high-energy effective action,''
{\it Phys. Rev. D}  {\bf 60} 014020 (1999).

 \bibitem{Kovchegov}
  Y. V. Kovchegov,
  %``Unitarization of the BFKL pomeron on a nucleus,''
{\it  Phys.\ Rev.\ D }{\bf 61}, 074018 (2000).
  %%CITATION = HEP-PH 9905214;%%  

 
  
\bibitem{Marquet:2016cgx}
  C. Marquet, E. Petreska and C. Roiesnel,
  %``Transverse-momentum-dependent gluon distributions from JIMWLK evolution,''
  {\it JHEP }{\bf 1610}, 065 (2016).  
  
\bibitem{Sun:2011iw} 
  P. Sun, B. W. Xiao and F. Yuan,
  %``Gluon Distribution Functions and Higgs Boson Production at Moderate Transverse Momentum,''
{\it  Phys.\ Rev.\ D }{\bf 84}, 094005 (2011).
  
\bibitem{Boer:2011kf} 
  D. Boer, W. J. den Dunnen, C. Pisano, M. Schlegel and W. Vogelsang,
  %``Linearly Polarized Gluons and the Higgs Transverse Momentum Distribution,''
  {\it Phys.\ Rev.\ Lett.\  }{\bf 108}, 032002 (2012).
  
\bibitem{Schafer:2012yx} 
  A. Schafer and J. Zhou,
  %``Higgs boson production in high energy proton-nucleus collisions,''
  {\it Phys.\ Rev.\ D }{\bf 85}, 114004 (2012).  
  
\bibitem{Kotko:2015ura}
  P. Kotko, K. Kutak, C. Marquet, E. Petreska, S. Sapeta and A. van Hameren,
  %``Improved TMD factorization for forward dijet production in dilute-dense hadronic collisions,''
  {\it JHEP }{\bf 1509} 106 (2015).
  
\bibitem{Marquet:2007vb} 
  C. Marquet,
  %``Forward inclusive dijet production and azimuthal correlations in p(A) collisions,''
  {\it Nucl.\ Phys.\ A }{\bf 796}, 41 (2007).
  
\bibitem{Iancu:2013dta} 
  E. Iancu and J. Laidet,
  %``Gluon splitting in a shockwave,''
 {\it Nucl.\ Phys.\ A }{\bf 916}, 48 (2013).    
 
\bibitem{Dumitru:2005gt} 
  A. Dumitru, A. Hayashigaki and J. Jalilian-Marian,
  %``The Color glass condensate and hadron production in the forward region,''
  {\it Nucl.\ Phys.\ A }{\bf 765}, 464 (2006). 
  
 
\bibitem{GolecBiernat:1998js} 
  K. J. Golec-Biernat and M. Wusthoff,
  %``Saturation effects in deep inelastic scattering at low Q**2 and its implications on diffraction,''
 {\it Phys.\ Rev.\ D }{\bf 59}, 014017 (1998). 
 
\bibitem{Sudakov:1954sw} 
  V.~V.~Sudakov,
  %``Vertex parts at very high-energies in quantum electrodynamics,''
 {\it Sov.\ Phys.\ JETP }{\bf 3}, 65 (1956)
  [{\it Zh.\ Eksp.\ Teor.\ Fiz.\  }{\bf 30}, 87 (1956)]. 
  
\bibitem{Mueller:2012uf} 
  A. H. Mueller, B. W. Xiao and F. Yuan,
  %``Sudakov Resummation in Small-$x$ Saturation Formalism,''
  {\it Phys.\ Rev.\ Lett.\  }{\bf 110}, no. 8, 082301 (2013).
  
\bibitem{Mueller:2013wwa} 
  A. H. Mueller, B. W. Xiao and F. Yuan,
  %``Sudakov double logarithms resummation in hard processes in the small-x saturation formalism,''
  {Phys.\ Rev.\ D }{\bf 88}, no. 11, 114010 (2013).   
  
\bibitem{Balitsky:2015qba} 
  I. Balitsky and A. Tarasov,
  %``Rapidity evolution of gluon TMD from low to moderate x,''
  {\it JHEP }{\bf 1510}, 017 (2015).
  
\bibitem{Balitsky:2016dgz} 
  I. Balitsky and A. Tarasov,
  %``Gluon TMD in particle production from low to moderate x,''
  {\it JHEP }{\bf 1606}, 164 (2016).   
 
\bibitem{Xiao:2017yya} 
  B. W. Xiao, F. Yuan and J. Zhou,
  %``Transverse Momentum Dependent Parton Distributions at Small-x,''
  {\it Nucl.\ Phys.\ B }{\bf 921}, 104 (2017).  
 
\bibitem{Catani:1990eg} 
  S. Catani, M. Ciafaloni and F. Hautmann,
  %``High-energy factorization and small x heavy flavor production,''
  {\it Nucl.\ Phys.\ B }{\bf 366}, 135 (1991).  
  
\bibitem{Deak:2009xt} 
  M. Deak, F. Hautmann, H. Jung and K. Kutak,
  %``Forward Jet Production at the Large Hadron Collider,''
  {\it JHEP }{\bf 0909}, 121 (2009).   
  
\bibitem{vanHameren:2016ftb} 
  A. van Hameren, P. Kotko, K. Kutak, C. Marquet, E. Petreska and S. Sapeta,
  %``Forward di-jet production in p+Pb collisions in the small-x improved TMD factorization framework,''
 {\it JHEP }{\bf 1612}, 034 (2016). 
 
\bibitem{Kotko:2017oxg} 
  P. Kotko, K. Kutak, S. Sapeta, A. M. Stasto and M. Strikman,
  %``Estimating nonlinear effects in forward dijet production in ultra-peripheral heavy ion collisions at the LHC,''
  {\it Eur.\ Phys.\ J.\ C }{\bf 77}, no. 5, 353 (2017).   
  
\bibitem{Kovchegov:2015zha} 
  Y. V. Kovchegov and M. D. Sievert,
  %``Calculating TMDs of a Large Nucleus: Quasi-Classical Approximation and Quantum Evolution,''
  {\it Nucl.\ Phys.\ B }{\bf 903}, 164 (2016).  
  
   \bibitem{Dumitru:2016jku} 
  A. Dumitru and V. Skokov,
  %``$cos(4φ$) azimuthal anisotropy in small-$x$ DIS dijet production beyond the leading power TMD limit,''
  {\it Phys.\ Rev.\ D }{\bf 94}, no. 1, 014030 (2016).
  
   \bibitem{Boer:2016fqd} 
  D. Boer, P. J. Mulders, C. Pisano and J. Zhou,
  %``Asymmetries in Heavy Quark Pair and Dijet Production at an EIC,''
  {\it JHEP }{\bf 1608}, 001 (2016).
  
  
  \bibitem{Boer:2016xqr} 
  D. Boer, S. Cotogno, T. van Daal, P. J. Mulders, A. Signori and Y. J. Zhou,
  %``Gluon and Wilson loop TMDs for hadrons of spin $\leq$ 1,''
  {\it JHEP }{\bf 1610}, 013 (2016).
  
     
  \bibitem{Szymanowski:2016mbq} 
  L. Szymanowski and J. Zhou,
  %``The spin dependent odderon in the diquark model,''
  {\it Phys.\ Lett.\ B }{\bf 760}, 249 (2016).
  
  \bibitem{Kovchegov:2016zex} 
  Y. V. Kovchegov, D. Pitonyak and M. D. Sievert,
  %``Helicity Evolution at Small $x$: Flavor Singlet and Non-Singlet Observables,''
  {\it Phys.\ Rev.\ D }{\bf 95}, no. 1, 014033 (2017).
  
  \bibitem{Kovchegov:2016weo} 
  Y. V. Kovchegov, D. Pitonyak and M. D. Sievert,
  %``Small-$x$ asymptotics of the quark helicity distribution,''
  {\it Phys.\ Rev.\ Lett.\  }{\bf 118}, no. 5, 052001 (2017).
  
   \bibitem{Marquet:2017xwy}
  C. Marquet, C. Roiesnel and P. Taels,
  %``Linearly polarized small-$x$ gluons in forward heavy-quark pair production,''
 {\it Phys.\ Rev.\ D }{\bf 97} no.1,  014004 (2018). 
 

\bibitem{Boer:2017xpy} 
  D. Boer, P. J. Mulders, J. Zhou and Y. j. Zhou,
  %``Suppression of maximal linear gluon polarization in angular asymmetries,''
  {\it JHEP }{\bf 1710}, 196 (2017).
  
   
\bibitem{Kovchegov:2017lsr} 
  Y.~V.~Kovchegov, D.~Pitonyak and M.~D.~Sievert,
  %``Small-$x$ Asymptotics of the Gluon Helicity Distribution,''
  JHEP {\bf 1710}, 198 (2017)
  
 \bibitem{Kovchegov:2017jxc} 
  Y. V. Kovchegov, D. Pitonyak and M. D. Sievert,
  %``Small-$x$ Asymptotics of the Quark Helicity Distribution: Analytic Results,''
  {\it Phys.\ Lett.\ B }{\bf 772}, 136 (2017).  
 
\bibitem{Boer:2003tx} 
  D. Boer and W. Vogelsang,
  %``Asymmetric jet correlations in p p uparrow scattering,''
  {\it Phys.\ Rev.\ D }{\bf 69}, 094025 (2004). 

\bibitem{Qiu:2007ar} 
  J. W. Qiu, W. Vogelsang and F. Yuan,
  %``Asymmetric di-jet production in polarized hadronic collisions,''
  {\it Phys.\ Lett.\ B }{\bf 650}, 373 (2007).
  
\bibitem{Qiu:2007ey} 
  J. W. Qiu, W. Vogelsang and F. Yuan,
  %``Single Transverse-Spin Asymmetry in Hadronic Dijet Production,''
  {\it Phys.\ Rev.\ D }{\bf 76}, 074029 (2007).
  
\bibitem{Collins:2007nk} 
  J. Collins and J. W. Qiu,
  %``$k_{T}$ factorization is violated in production of high-transverse-momentum particles in hadron-hadron collisions,''
  {\it Phys.\ Rev.\ D }{\bf 75}, 114014 (2007).
  
\bibitem{Rogers:2010dm} 
  T. C. Rogers and P. J. Mulders,
  %``No Generalized TMD-Factorization in Hadro-Production of High Transverse Momentum Hadrons,''
  {\it Phys.\ Rev.\ D }{\bf 81}, 094006 (2010).
  
\bibitem{Xiao:2010sp} 
  B. W. Xiao and F. Yuan,
  %``Non-Universality of Transverse Momentum Dependent Parton Distributions at Small-x,''
  {\it Phys.\ Rev.\ Lett.\  }{\bf 105}, 062001 (2010).  
  
  

 
\bibitem{Ji:2002aa} 
  X. d. Ji and F. Yuan,
  %``Parton distributions in light cone gauge: Where are the final state interactions?,''
  {\it Phys.\ Lett.\ B }{\bf 543}, 66 (2002). 
  
\bibitem{8}
  D. Boer, P. J. Mulders and F. Pijlman,
  %``Universality of T odd effects in single spin and azimuthal asymmetries,''
 {\it  Nucl.\ Phys.\ B } {\bf 667} 201 (2003).
%  [hep-ph/0303034].
  %%CITATION = HEP-PH/0303034;%%  
  
\bibitem{JalilianMarian:1996xn} 
  J. Jalilian-Marian, A. Kovner, L. D. McLerran and H. Weigert,
  %``The Intrinsic glue distribution at very small x,''
  {\it Phys.\ Rev.\ D }{\bf 55}, 5414 (1997).
  
\bibitem{Kovchegov:1998bi} 
  Y. V. Kovchegov and A. H. Mueller,
  %``Gluon production in current nucleus and nucleon - nucleus collisions in a quasiclassical approximation,''
  {\it Nucl.\ Phys.\ B }{\bf 529}, 451 (1998).    
 
\bibitem{Blaizot:2004wu} 
  J. P. Blaizot, F. Gelis and R. Venugopalan,
  %``High-energy pA collisions in the color glass condensate approach. 1. Gluon production and the Cronin effect,''
  {\it Nucl.\ Phys.\ A }{\bf 743}, 13 (2004). 
 

 
\bibitem{Gelis:2002ki} 
  F. Gelis and J. Jalilian-Marian,
  %``Photon production in high-energy proton nucleus collisions,''
  {\it Phys.\ Rev.\ D }{\bf 66}, 014021 (2002). 
 

 
\bibitem{Altinoluk:2011qy} 
  T. Altinoluk and A. Kovner,
  %``Particle Production at High Energy and Large Transverse Momentum - 'The Hybrid Formalism' Revisited,''
  {\it Phys.\ Rev.\ D }{\bf 83}, 105004 (2011).
  
\bibitem{Marquet:2009ca} 
  C. Marquet, B. W. Xiao and F. Yuan,
  %``Semi-inclusive Deep Inelastic Scattering at small x,''
 {\it Phys.\ Lett.\ B }{\bf 682}, 207 (2009).   
  
\bibitem{Akcakaya:2012si} 
  E. Akcakaya, A. Schäfer and J. Zhou,
  %``Azimuthal asymmetries for quark pair production in pA collisions,''
  {\it Phys.\ Rev.\ D }{\bf 87}, no. 5, 054010 (2013).
  
 \bibitem{Benic:2017znu} 
  S. Benic and A. Dumitru,
  %``Prompt photon - jet angular correlations at central rapidities in p+A collisions,''
  {\it Phys.\ Rev.\ D }{\bf 97}, no. 1, 014012 (2018).  
  
 
  
\bibitem{Altinoluk:2018uax}
  T. Altinoluk, N. Armesto, A. Kovner, M. Lublinsky and E. Petreska,
  %``Soft photon and two hard jets forward production in proton-nucleus collisions,''
  arXiv:1802.01398 [hep-ph].  
  
 
  
   
  
  

\bibitem{Boer:2016bfj} 
  D. Boer,
  %``Gluon TMDs in quarkonium production,''
 {\it Few Body Syst.\  }{\bf 58}, no. 2, 32 (2017). 
 
\bibitem{Blaizot:2002np} 
  J. P. Blaizot, E. Iancu and H. Weigert,
  %``Nonlinear gluon evolution in path integral form,''
  {\it Nucl.\ Phys.\ A }{\bf 713}, 441 (2003). 
 
\bibitem{Rummukainen:2003ns} 
  K. Rummukainen and H. Weigert,
  %``Universal features of JIMWLK and BK evolution at small x,''
  {\it Nucl.\ Phys.\ A }{\bf 739}, 183 (2004). 
  
\bibitem{Dumitru:2011vk} 
  A. Dumitru, J. Jalilian-Marian, T. Lappi, B. Schenke and R. Venugopalan,
  %``Renormalization group evolution of multi-gluon correlators in high energy QCD,''
  {\it Phys.\ Lett.\ B }{\bf 706}, 219 (2011).
  
\bibitem{Lappi:2012vw} 
  T. Lappi and H. Mäntysaari,
  %``On the running coupling in the JIMWLK equation,''
 {\it Eur.\ Phys.\ J.\ C }{\bf 73}, no. 2, 2307 (2013).
 
\bibitem{Dumitru:2015gaa} 
  A. Dumitru, T. Lappi and V. Skokov,
  %``Distribution of Linearly Polarized Gluons and Elliptic Azimuthal Anisotropy in Deep Inelastic Scattering Dijet Production at High Energy,''
  {\it Phys.\ Rev.\ Lett.\  }{\bf 115}, no. 25, 252301 (2015).     

\bibitem{Stasto:2000er} 
  A. M. Stasto, K. J. Golec-Biernat and J. Kwiecinski,
  %``Geometric scaling for the total gamma* p cross-section in the low x region,''
 {\it Phys.\ Rev.\ Lett.\  }{\bf 86}, 596 (2001).
    
\bibitem{Munier:2003vc} 
  S. Munier and R. B. Peschanski,
  %``Geometric scaling as traveling waves,''
  {\it Phys.\ Rev.\ Lett.\  }{\bf 91}, 232001 (2003).    
  
\bibitem{Dominguez:2011gc} 
  F. Dominguez, A. H. Mueller, S. Munier and B. W. Xiao,
  %``On the small-$x$ evolution of the color quadrupole and the Weizsäcker–Williams gluon distribution,''
 {\it Phys.\ Lett.\ B }{\bf 705}, 106 (2011).
   
  
\bibitem{Lappi:2011ju} 
  T. Lappi,
  %``Gluon spectrum in the glasma from JIMWLK evolution,''
 { Phys.\ Lett.\ B }{\bf 703}, 325 (2011).  
 
\bibitem{Adare:2011sc} 
  A. Adare {\it et al.} [PHENIX Collaboration],
  %``Suppression of back-to-back hadron pairs at forward rapidity in $d+$Au Collisions at $\sqrt{s_{NN}}=200$ GeV,''
  {\it Phys.\ Rev.\ Lett.\  }{\bf 107}, 172301 (2011).
  
\bibitem{Braidot:2010zh} 
  E. Braidot [STAR Collaboration],
  %``Suppression of Forward Pion Correlations in d+Au Interactions at STAR,''
  arXiv:1005.2378 [hep-ph]. 
  
 
\bibitem{Albacete:2010pg} 
  J. L. Albacete and C. Marquet,
  %``Azimuthal correlations of forward di-hadrons in d+Au collisions at RHIC in the Color Glass Condensate,''
 {\it Phys.\ Rev.\ Lett.\  }{\bf 105}, 162301 (2010).  
 
\bibitem{Stasto:2011ru} 
  A. Stasto, B. W. Xiao and F. Yuan,
  %``Back-to-Back Correlations of Di-hadrons in dAu Collisions at RHIC,''
  {\it Phys.\ Lett.\ B }{\bf 716}, 430 (2012). 
  
\bibitem{Lappi:2012nh}
  T. Lappi and H. Mantysaari,
  %``Forward dihadron correlations in deuteron-gold collisions with the Gaussian approximation of JIMWLK,''
  {\it Nucl.\ Phys.\ A }{\bf 908} 51 (2013).
%  [arXiv:1209.2853 [hep-ph]].
  %%CITATION = ARXIV:1209.2853;%%
  
\bibitem{Qiu:2004da} 
  J. w. Qiu and I. Vitev,
  %``Coherent QCD multiple scattering in proton-nucleus collisions,''
  {\it Phys.\ Lett.\ B }{\bf 632}, 507 (2006).
  
 \bibitem{Kang:2011bp} 
  Z. B. Kang, I. Vitev and H. Xing,
  %``Dihadron momentum imbalance and correlations in d+Au collisions,''
 {\it Phys.\ Rev.\ D }{\bf 85}, 054024 (2012).   
  
 \bibitem{Zheng:2014vka} 
  L. Zheng, E. C. Aschenauer, J. H. Lee and B. W. Xiao,
  %``Probing Gluon Saturation through Dihadron Correlations at an Electron-Ion Collider,''
 {\it Phys.\ Rev.\ D }{\bf 89}, no. 7, 074037 (2014). 
  
  

  
\bibitem{vanHameren:2014ala}
  A. van Hameren, P. Kotko, K. Kutak and S. Sapeta,
  %``Small-$x$ dynamics in forward-central dijet decorrelations at the LHC,''
 {\it Phys.\ Lett.\ B }{\bf 737} 335 (2014).  
 
\bibitem{Kutak:2012rf}
  K. Kutak and S. Sapeta,
  %``Gluon saturation in dijet production in p-Pb collisions at Large Hadron Collider,''
  {\it Phys.\ Rev.\ D {\bf 86} } 094043 (2012).
%  [arXiv:1205.5035 [hep-ph]].
  %%CITATION = ARXIV:1205.5035;%% 
  
\bibitem{vanHameren:2014lna}
  A. van Hameren, P. Kotko, K. Kutak, C. Marquet and S. Sapeta,
  %``Saturation effects in forward-forward dijet production in p+Pb
  %collisions,''
  {\it Phys.\ Rev.\ D }{\bf 89} 094014 (2014).
%  [arXiv:1402.5065 [hep-ph]].
  %%CITATION = ARXIV:1402.5065;%%

\bibitem{vanHameren:2013fla}
  A. van Hameren, P. Kotko and K. Kutak,
  %``Three jet production and gluon saturation effects in p-p and p-Pb collisions within high-energy factorization,''
  {\it Phys.\ Rev.\ D } {\bf 88}  9,  094001 (2013)
   [Erratum-ibid.\ D {\bf 90}  3,  039901 (2014)].
%  [arXiv:1308.0452 [hep-ph]].
  %%CITATION = ARXIV:1308.0452;%%  
 
\bibitem{Dumitru:2002qt} 
  A. Dumitru and J. Jalilian-Marian,
  %``Forward quark jets from protons shattering the colored glass,''
 {\it Phys.\ Rev.\ Lett.\  }{\bf 89}, 022301 (2002). 
 
\bibitem{Benic:2016uku} 
  S. Benic, K. Fukushima, O. Garcia-Montero and R. Venugopalan,
  %``Probing gluon saturation with next-to-leading order photon production at central rapidities in proton-nucleus collisions,''
  {\it JHEP }{\bf 1701}, 115 (2017). 
  
\bibitem{Levin:2010zs} 
  E. Levin,
  %``Gluon saturation and inclusive production at low transverse momenta,''
  {\it Phys.\ Rev.\ D }{\bf 82}, 101704 (2010).  
  
\bibitem{Watanabe:2015yca} 
  K. Watanabe and B. W. Xiao,
  %``Forward Heavy Quarkonium Productions at the LHC,''
 {\it Phys.\ Rev.\ D }{\bf 92}, no. 11, 111502 (2015).  
  

 
    
%\bibitem{4} M. Tinkham, {\it Group Theory and Quantum Mechanics}
%(McGraw-Hill, New York, 1964).



%\bibitem{4} T. Tel, in {\it Experimental Study and Characterization of
%Chaos}, ed. B. Hao (World Scientific, Singapore, 1990),
%p. 149.
%
%\bibitem{5}
%P. P. Edwards, in {\it Superconductivity and Applications
%--- Proc. Taiwan Int. Symp. on Superconductivity}, ed.  P. T. Wu
%{\it et al.} (World Scientific, Singapore, 1989), p. 29.
%
%\bibitem{6}
%W. J. Johnson, Ph.D. Thesis, Univ. of Wisconsin, Madison (1968).
%
%\bibitem{7}
%P. F. Marteau and H. D. I. Arbabanel, ``Noise reduction in
%chaotic time series using scaled probabilistic methods'',
%UCSD/INLS preprint, October 1990.


\end{thebibliography}
\end{document}